\documentclass[12pt,preprint]{aastex}

\newcommand{\kakkoi}[3]{\left(\frac{{#1}}{{#2}}\right)^{#3}}
\newcommand{\BF}[1]{\mbox{\normalsize \boldmath $#1$}}
\newcommand{\e}[1]{\times 10^{#1}}
\newcommand{\ave}[1]{\langle #1 \rangle}

\renewcommand{\apj}{Astrophys. J.\,\,}
\renewcommand{\aap}{Astron. Astrophys.\,\,}
\renewcommand{\icarus}{Icarus\,\,}
\usepackage{amsmath}
\usepackage{color}

\shorttitle{Accretion Rates of Planetesimals in Nebular Gas}
\shortauthors{Tanigawa \& Ohtsuki}

\begin{document}


\title{Accretion Rates of Planetesimals by Protoplanets Embedded in
Nebular Gas}



\author{Takayuki Tanigawa}
\affil{Department of Earth and Planetary Sciences, Tokyo Institute of
Technology, Ookayama, Meguro-ku, Tokyo 152-8551, Japan, \\
tel: +81-3-5734-2339, fax: +81-3-5734-3538;\\
and Laboratory for Atmospheric and Space Physics, University of
Colorado, 392 UCB, Boulder, Colorado 80309-0392, USA;}
\email{tanigawa@geo.titech.ac.jp}

\and

\author{Keiji Ohtsuki}

\affil{Laboratory for Atmospheric and Space Physics,
University of Colorado, 392 UCB, Boulder, Colorado 80309-0392, USA; //
and
Department of Earth and Planetary Sciences and Center for Planetary
Science, Kobe University, Kobe 657-8501, Japan}
\email{ohtsuki@tiger.kobe-u.ac.jp}





\begin{abstract}
When protoplanets growing by accretion of planetesimals have
atmospheres, small planetesimals approaching the protoplanets lose their
energy by gas drag from the atmospheres, which leads them to be captured
within the Hill sphere of the protoplanets.  As a result, growth rates
of the protoplanets are enhanced.
In order to study the effect of an atmosphere on planetary growth rates,
we performed numerical integration of orbits of planetesimals for a wide
range of orbital elements and obtained the effective accretion rates of
planetesimals onto planets that have atmospheres.
Numerical results are obtained as a function of planetesimals'
eccentricity, inclination, planet's radius, and non-dimensional gas-drag
parameters which can be expressed by several physical quantities such as
the radius of planetesimals and the mass of the protoplanet.
Assuming that the radial distribution of the gas density near the
surface can be approximated by a power-law, we performed analytic
calculation for the loss of planetesimals' kinetic energy due to gas
drag, and confirmed agreement with numerical results.
We confirmed that the above approximation of the power-law density
distribution is reasonable for accretion rate of protoplanets with one
to ten Earth-masses, unless the size of planetesimals is too small.
We also calculated the accretion rates of planetesimals averaged over a
Rayleigh distribution of eccentricities and inclinations, and derived a
semi-analytical formula of accretion rates, which reproduces the
numerical results very well.
Using the obtained expression of the accretion rate, we examined the
growth of protoplanets in nebular gas.  We found that the effect of
atmospheric gas drag can enhance the growth rate significantly,
depending on the size of planetesimals.
\end{abstract}






\section{Introduction}
Planets are thought to be formed in circum-stellar gas disks by
accumulating a large number of planetesimals.  When a protoplanet
reaches about the Moon mass, it starts to attract surrounding gas of the
disk by its gravity, which would lead to the formation of a thick
atmosphere, if the growth of the protoplanet proceeds in the nebular
gas.  Thus, in the process of planetesimal accretion onto planets,
interactions between planetesimals and atmospheres should be considered.
Interactions between planetary atmospheres and planetesimals have been
studied mainly in the context of giant planet formation.
\citet{Podolak1988} studied the interaction of planetesimals with
atmospheres in order to examine the ablation and dissolution of incoming
planetesimals due to gas drag by calculating trajectories and mass-loss
rates of the planetesimals.  \citet{Pollack1996} and
\citet{Hubickyj2005} followed the method of \citet{Podolak1988} to
calculate energy and mass deposition on the atmosphere when
planetesimals are passing through it as a part of their formation model
of giant planets.  Also, in relation to the origin of natural
satellites, the interaction between primordial atmospheres and incoming
objects was considered, in order to study the possibility of capture of
satellites by the gas drag, as well as the orbital evolution of captured
satellites \citep{Pollack1979, Cuk2004}.

Interactions between planetesimals and atmospheres also affect the
growth rate of protoplanets.  In protoplanetary disks, solid
protoplanets form first, and then solid protoplanets with a sufficiently
large mass (about $\gtrsim$ 10 Earth masses) acquire a large amount of
gas to become gas-giant protoplanets \citep{Mizuno1980, Bodenheimer1986,
Pollack1996, Ikoma2000, Hubickyj2005}.  This scenario is widely
accepted, but the growth timescale of solid planets predicted by results
of theoretical studies of planetary accretion
\citep[e.g.,][]{Tanaka1999,KokuboIda2000} are too long in two senses.
First, the predicted growth timescale is longer than the typical
timescale of inward migration of protoplanets predicted by current
theories including linear analyses \citep[e.g.,][]{Ward1986, Ward1997,
Korycansky1993, Tanaka2002} and numerical simulations
\citep[e.g.,][]{DAngelo2002,Masset2006}, which results in the loss of
forming planets into the central star before the completion of their
growth.
Second, planets cannot reach a critical mass that is necessary for the
onset of runaway gas accretion before dissipation of disk gas, which
means that gas giant planets are difficult to be formed.
%
Current theories predict that solid planets grow in proto-planetary
disks via runaway growth phase \citep{WS89} followed by oligarchic
growth phase \citep{KokuboIda1996}, and in both phases, protoplanets
grow mainly by capturing objects that are much smaller than the
protoplanets.
Thus, in order to estimate the growth timescale of solid planets, one
has to investigate the stage where a large number of small objects are
accreted by a small number of large objects.

Collision rates of small objects onto protoplanets have been studied in
detail by numerical orbital integration and analytical calculations
\citep{Nishida1983, Wetherill1985, IN89, GL90, GL92, Dones1993}.
The early stage of planetary accretion is likely to have proceeded in
the presence of the protoplanetary gas disk, but most of the previous
studies did not consider the effect of atmosphere around protoplanets,
which likely enhances effective accretion rates through gas drag.
\citet{Kary1993} and \citet{Kary1995} studied accretion rates when
planetesimals are migrating radially inward by gas drag from the nebular
gas, but they did not consider the effect of planets' atmospheres.

\citet{II03} first examined the effect of atmospheric gas drag on
planetary accretion rates in detail.  They showed that the presence of
atmospheres around planets can enhance the accretion rate greatly
depending on the size of objects that are passing through the
atmospheres.
The results were used in their simulation of planetary accretion in the
Jovian region \citep{Inaba_etal_2003}.
\citet{Chambers2006a,Chambers2006b} also used the results of
\citet{II03} in his semi-analytic model of oligarchic growth of planets,
and showed that the enhancement of the accretion rate can shorten the
growth timescale so that gas giant planets can be formed before the
dispersal of the gas disk, if incoming planetesimals are small
($\lesssim$ 100m).
Although \citet{II03} demonstrated the importance of the atmosphere for
planetary accretion rates, most of their results were based on
calculations neglecting the solar gravity, and three-body orbital
integration was performed only in the case of initially circular orbits.
In addition, they used a realistic model for the atmospheric structure
rather than a simplified one, which makes the study of
parameter-dependence or comparison with analytic calculations rather
difficult.

In the present work, we obtain accretion rates of planetesimals onto
protoplanets that have atmospheres, by using three-body orbital
integration for a wide range of parameters.
We use a simple power-law density distribution model for the atmosphere,
which allows a systematic study of the accretion rate and its
parameter-dependence.
From the obtained accretion rates, we derive a semi-analytical
expression of the accretion rates for planets with atmospheres and
examine growth of protoplanets in the nebular gas.
In \S \ref{sec_setup}, we first show the basic formulation and some
analytic estimations which are the basis of our numerical simulation
and analysis to obtain the accretion rates.
In \S \ref{sec_numerical_simulation}, we show the results of our
numerical simulations to obtain the accretion rates, together with some
demonstrative calculations, which are useful in understanding the basic
behavior of planetesimals' motion under the gas drag of planetary
atmospheres.
In \S \ref{sec_empirical_formula}, we derive a semi-analytic formula for
accretion rates under the assumption of power-law density profile for
atmospheres.
In \S \ref{sec_realistic_atmosphere}, we examine the validity of the
power-law approximation for density profile of atmospheres.
Using the obtained semi-analytic formula for accretion rates, we examine
the growth of protoplanets with atmospheres in \S \ref{sec_growth}.
Our conclusions and discussion are presented in \S \ref{sec_summary}.

\section{Basic Formulation and Analytical Estimate}\label{sec_setup}
\subsection{Basic equations}
When the masses of planetesimals and the planet are much smaller than
the solar mass and their orbital eccentricities and inclinations are
sufficiently small, their motions in a rotating coordinate system can
be described by linearized equations, called Hill's equations.  We use a
coordinate system centered on the planet, where the $x$-axis points
radially outwards, the $y$-axis points in the direction of the planet's
orbital motion, and the $z$-axis is normal to the $x$--$y$ plane.  We
scale the distance by the mutual Hill radius $r_{\rm H} \equiv ah$ with
$h \equiv \{(M+m)/3M_\ast\}^{1/3}$ ($M$ and $m$ are the masses of the
planet and a planetesimal, respectively; $a$ is the semi-major axis of
the planet; and $M_\ast$ is the mass of the central star) and the time
by $\Omega_{\rm K}^{-1}$ ($\Omega_{\rm K}$ is the planet's orbital
angular frequency).  The non-dimensional equation for the relative
motion between the planet and a planetesimal can then be written as
\begin{equation}
\frac{d\tilde{\BF{v}}}{d\tilde{t}}
 = - \nabla \tilde{\Phi}
   - 2 \BF{e}_z \times \tilde{\BF{v}}
   + \tilde{\BF{a}}_{\rm drag},
\label{EOM}
\end{equation}
where $\tilde{\BF{v}}$ is velocity, $\tilde{t}$ is time, and
$\tilde{\Phi}$ is the Hill potential given by
\begin{equation}
\tilde{\Phi}
 = - \frac{3}{\tilde{r}}
   - \frac{3}{2} \tilde{x}^2
   + \frac{1}{2} \tilde{z}^2
   + \frac{9}{2},
\end{equation}
where $\tilde{r} = \sqrt{\tilde{x}^2 + \tilde{y}^2 + \tilde{z}^2}$.
Variables with tildes denote non-dimensional quantities.  The
acceleration due to gas drag $\tilde{\BF{a}}_{\rm drag}$ is described by
\begin{equation}
\tilde{\BF{a}}_{\rm drag}
 \equiv
    \frac{\BF{F}_{\rm drag}/m}{r_{\rm H}\Omega_{\rm K}^2}
 = - \frac{3}{8} C_{\rm D}
     \frac{\rho_{\rm g}}{\rho_{\rm s}}
     \tilde{r}_{\rm s}^{-1} \Delta \tilde{u} \Delta \tilde{\BF{u}},
\label{a_drag_define}
\end{equation}
where $\BF{F}_{\rm drag} = - (C_{\rm D}/2)\pi r_{\rm s}^2 \rho_{\rm g}
\Delta u \Delta \BF{u}$ is the drag force for a planetesimal with radius
$r_{\rm s}$, $m$ is the mass of the planetesimal, $C_{\rm D}$ is the gas
drag coefficient of order unity, $\rho_{\rm g}$ is gas density,
$\rho_{\rm s}$ is the internal density of planetesimals, $\tilde{r}_{\rm
s}$ is the normalized physical radius of the planetesimals, and $\Delta
\BF{u}$ is the velocity of the objects relative to the gas.

\subsection{Atmospheric Structure}
In the present work, we assume that the density structure of the
atmosphere is spherically symmetric.  For sufficiently massive solid
protoplanets ($>$ 10 Earth masses), runaway gas accretion occurs
and the density distribution of the inflowing gas significantly deviates
from spherical symmetry.  However, spherically symmetric atmosphere is a
good approximation for atmospheres of low-mass protoplanets because
hydrostatic (or quasi-hydrostatic) equilibrium is realized in such cases
\citep[e.g.,][]{Mizuno1980,Bodenheimer1986,Pollack1996,Ikoma2000}.
In addition, in order to study the gas drag effect systematically, we
assume a power-law function for the atmospheric structure as
\begin{equation}
\rho_{\rm g}
 = \rho_{\rm b} \kakkoi{\tilde{r}}{\tilde{r}_{\rm p}}{-\alpha},
\label{rho_g}
\end{equation}
where $\rho_{\rm b}$ is the density at the bottom of the atmosphere, or
equivalently, the density at the surface of the solid core, $\tilde{r}$
is the distance from the center of the planet, $\tilde{r}_{\rm p}$ is
the radius of the solid part of the planet, and the exponent $\alpha$ is
assumed to be constant.

For a purely radiative atmosphere with a constant opacity, the density
distribution is given by the above power-law with $\alpha=3$ (e.g.,
\citet{Stevenson1982}; see Appendix A).
In more realistic atmospheric structures, on the other hand, the slope
of the density distribution changes near the sublimation points of ice
and silicate dust, especially when opacity (or dust/gas ratio) is large
\citep[e.g.,][]{II03,Rafikov2006}, and outside of Bondi radius where gas
is no longer bounded by the planet gravity.  However, the overall
structure in most radial regions of atmospheres can still be well
approximated by a power-law distribution.
In the present work, we adopt the above simple model for atmospheric
structures in order to examine the effect of atmosphere systematically
for a wide range of parameters and to better understand physics behind
it.  Effects of a more realistic atmospheric structure based on the
analytic solution described in \citet{II03} will be discussed in Section
\ref{sec_realistic_atmosphere}.
Once a power-law function is assumed for gas density, effects of gas
drag can be described by two parameters: the exponent $\alpha$, and the
non-dimensional gas drag parameter $\xi$, which will be defined in the
next subsection.

\subsection{Non-dimensional gas drag parameter $\xi$}
Assuming that density profile is expressed by Eq.~(\ref{rho_g}),
Eq.~(\ref{a_drag_define}) can be rewritten in the form
\begin{equation}
\tilde{\BF{a}}_{\rm drag}
 = -\xi \tilde{r}^{-\alpha} \Delta \tilde{u} \Delta \tilde{\BF{u}},
\end{equation}
where $\xi$ is a non-dimensional parameter representing the strength of
gas drag, and is defined by
\begin{eqnarray}
\xi
&\equiv&
  \frac{3}{8} C_{\rm D}
  \tilde{r}_{\rm s}^{-1}
  \frac{\rho_{\rm b}}{\rho_{\rm s}}
  \tilde{r}_{\rm p}^\alpha, \nonumber \\
&=&4.3\e{-5}
   \kakkoi{a}{\rm 1AU}{-2}
   \kakkoi{M}{M_\oplus}{1/3}
   \kakkoi{r_{\rm s}}{\rm 1km}{-1}
   \kakkoi{\rho_{\rm b}}{\rm 1kg\,m^{-3}}{}
   \kakkoi{\rho_{\rm s}}{\rm 10^3kg\,m^{-3}}{-1} \nonumber \\
& &\times
   \kakkoi{\rho_{\rm core}}{\rho_\oplus}{-1}
   \kakkoi{M_\ast}{M_\odot}{2/3}
   \kakkoi{C_{\rm D}}{1}{}.
\label{xi_tmp}
\end{eqnarray}
In the above, $M_\oplus$ is the Earth mass, $\rho_{\rm core}$ and
$\rho_\oplus$ are the internal density of planets and the Earth,
respectively, and we assumed $\alpha=3$ as mentioned above.
For a purely radiative atmosphere with large optical thickness,
$\rho_{\rm b} \propto M^3 L^{-1} f_\kappa^{-1}$ is a good approximation
(see Appendix A or \citet{Sasaki1990}), where $L$ is the
released energy at the solid surface of the planet per unit time due to
accretion of incoming planetesimals, and $f_\kappa$ is the opacity
depletion factor, that is, $f_\kappa=1$ when dust to gas ratio is the
same as the ratio in interstellar clouds assuming dust optical property
is the same as that of interstellar dust; $f_\kappa$ decreases when dust
itself is depleted or when dust opacity decreases (due to growth of
dust, for example).
Thus $\rho_{\rm b}$, which is a function of the planet mass, can be
eliminated from the above equation.  Using the analytic formula for
atmospheric structures given by \citet{II03}, we found that $\rho_{\rm
b}$ can be approximated by $\rho_{\rm b} \simeq 0.17 {\rm kg\,m^{-3}}
(M/M_\oplus)^3 (L/10^{-7}L_\odot)^{-1}$ $(\rho_{\rm core}/\rho_\oplus)$
$(f_\kappa/0.01)^{-1}$ ($L_\odot$ is the solar luminosity) when
$f_\kappa$ is around 0.01.  Substitution of this $\rho_{\rm b}$ into
Eq.~(\ref{xi_tmp}) yields
\begin{eqnarray}
\xi
&=& 2.4\e{-4}
    \kakkoi{a}{\rm 1AU}{-2}
    \kakkoi{M}{M_\oplus}{8/3}
    \kakkoi{\dot{M}}{10^{-7}M_\oplus/\rm yr}{-1}
    \kakkoi{r_{\rm s}}{\rm 1km}{-1}
    \kakkoi{\rho_{\rm s}}{10^3 \rm kg\,m^{-3}}{-1} \nonumber \\
& & \times
    \kakkoi{\rho_{\rm core}}{\rho_\oplus}{-1/3}
    \kakkoi{M_\ast}{M_\odot}{2/3}
    \kakkoi{C_{\rm D}}{1}{}
    \kakkoi{f_\kappa}{0.01}{-1},
\label{xi_typical}
\end{eqnarray}
where we used accretion energy $L = GM\dot{M}/r_{\rm p}$ and $\dot{M}$
is the accretion rate of planetesimals onto the planet.

\subsection{Normalized accretion rate}
Following the definition of collision rate given by
\citet{Nakazawa1989}, we define the accretion rate of planetesimals per
unit surface number density as
\begin{equation}
P_{\rm acc}(\tilde{e},\tilde{i})
 = \int
        p_{\rm acc}(\tilde{e},\tilde{i},\tilde{b})
        \frac{3}{2}|\tilde{b}|
        d\tilde{b},
\label{Pacc}
\end{equation}
where $\tilde{e}$ and $\tilde{i}$ are planetesimals' orbital
eccentricity and inclination divided by $h$, and $\tilde{b}$ is the
initial value of the normalized semimajor axis of an incoming
planetesimal relative to the planet at $\tilde{y}=\infty$.  $p_{\rm
acc}(\tilde{e},\tilde{i},\tilde{b})$ is the accretion probability for a
given set of the orbital elements $(\tilde{e},\tilde{i},\tilde{b})$
defined by
\begin{equation}
p_{\rm acc}(\tilde{e},\tilde{i},\tilde{b})
 = \iint \varphi_{\rm acc}(\tilde{e},\tilde{i},\tilde{b},\tau,\varpi)
         \frac{d\tau d\varpi}{(2\pi)^2},
\end{equation}
where $\tau$ and $\varpi$ are the initial phase angles for epicyclic
motion in horizontal and vertical directions, respectively, and
$\varphi_{\rm acc}(\tilde{e},\tilde{i},\tilde{b},\tau,\varpi)$ is a
judgment function whether an object is accreted by the planet: unity if
the object is accreted, and zero otherwise.
In general, accretion can happen in two ways: (i) a planetesimal hits
the planet regardless of whether it loses enough energy to become bound,
or (ii) a planetesimal loses enough energy through gas drag to become
gravitationally bound.
We thus define the following two quantities in a similar way: $P_{\rm
col}$ is the normalized accretion rate due to direct collision with the
solid surface of a planet without an atmosphere, which is the definition
identical to that of \citet{Nakazawa1989}; and $P_{\rm cap}$ is the
normalized rate of capture of planetesimals within the Hill sphere of
the planet due to gas drag, with the assumption that the planet is a
point mass (i.e., direct collision onto the planet is neglected).  We
also define $p_{\rm col}$ and $p_{\rm cap}$ in a similar way.

When a planet does not have an atmosphere and the random velocity of
planetesimals is large enough, the rate of direct collision onto the
solid surface of the planet with radius $\tilde{r}_{\rm p}$ is
analytically given by \citep{GL90}
\begin{eqnarray}
P_{\rm col}
&=& \int_{-\tilde{e}}^{\tilde{e}}
       p_{\rm col} (\tilde{e},\tilde{i},\tilde{b})
       \frac{3}{2} |b| d\tilde{b} \nonumber \\
&=& \frac{2\tilde{r}_{\rm p}^2}{\pi \tilde{i}}
    \sqrt{\tilde{e}^2 + \tilde{i}^2}
    E(k)
    \left( 1 + \frac{6/\tilde{r}_{\rm p}}{\tilde{e}^2+\tilde{i}^2}
                \frac{K(k)}{E(k)}
    \right),
\label{Pcol}
\end{eqnarray}
where $K$ and $E$ are the complete elliptic integrals of the first and
second kinds, respectively, with $k \equiv
\sqrt{3\tilde{e}^2/4(\tilde{e}^2+\tilde{i}^2)}$.  Collision probability
$p_{\rm col}$ can be written as
\begin{equation}
p_{\rm col}(\tilde{e},\tilde{i},\tilde{b})
 = \frac{2\tilde{r}_{\rm p}^2}{3\pi |\tilde{b}| \tilde{i}}
   \sqrt{\frac{\tilde{e}^2+\tilde{i}^2-(3/4)\tilde{b}^2}
              {\tilde{e}^2-\tilde{b}^2}}
   \left( 1 + \frac{6/\tilde{r}_{\rm p}}
                   {\tilde{e}^2+\tilde{i}^2-(3/4)\tilde{b}^2}
   \right).
\label{integrand_of_pcol}
\end{equation}

\subsection{Analytic estimate of accretion rates due to atmospheric
  drag}
\label{sec_R_c}
Without energy dissipation, objects coming into the Hill sphere always
escape from the sphere in the end unless they collide with the planet,
but they can be captured within the Hill sphere if there is energy
dissipation such as atmospheric gas drag.  \citet{II03} introduced
``enhanced radius'' of a planet due to atmospheric gas drag;
planetesimals which come closer than this radius lose a significant
amount of kinetic energy and cannot escape out of the Hill radius, thus
it can be regarded as an effective radius of the planet for capturing
planetesimals.  In the present study, we assume a power-law density
distribution for atmosphere, which enables us to obtain the expression
for the enhanced radius analytically.

We first estimate energy dissipation of objects that pass through the
atmosphere in a way similar to \citet{II03}.
We assume that most of energy dissipation in one encounter occurs near
the point of closest approach to the planet, because the gas density
increases steeply with decreasing distance from the planet.  We will
confirm the validity of this assumption later, using orbital integration
(\S \ref{sec_gas_drag}).
Energy dissipation due to gas drag $\Delta \tilde{E}$ is then estimated
by the work acting on a unit mass of the object:
\begin{eqnarray}
\Delta \tilde{E}
&\sim& \tilde{a}_{\rm drag} \Delta \tilde{\ell} \nonumber \\
&=& f_{\rm c} \xi \tilde{r}_{\rm min}^{1-\alpha}
    \left( \tilde{v}_\infty^2
          +\frac{6}{\tilde{r}_{\rm min}}\right),
\label{Delta-E}
\end{eqnarray}
where $\tilde{a}_{\rm drag} = |\tilde{\BF{a}}_{\rm drag}|$; $\Delta
\tilde{\ell}$ is the path length of the orbit near the point of closest
approach and is substituted by $f_{\rm c} \tilde{r}_{\rm min}$ ($f_{\rm
c}$ is a correction factor of order unity, and $\tilde{r}_{\rm min}$ is
the minimum distance from the origin); and $\tilde{v}_\infty$ is the
relative velocity between the planet and an approaching planetesimal
before acceleration due to the planet's gravity.  When the relative
velocity is dominated by the random velocity rather than Kepler shear,
$\tilde{v}_\infty$ can be approximated by the relative velocity at the
origin in the unperturbed, non-gravitating solution to the Hill's
equation \citep{Nakazawa1989} given as
\begin{equation}
\tilde{v}_\infty
 = \sqrt{ \tilde{e}^2 + \tilde{i}^2 - \frac{3}{4}\tilde{b}^2}.
\label{v_infty}
\end{equation}

\begin{figure}
\epsscale{0.4}
\plotone{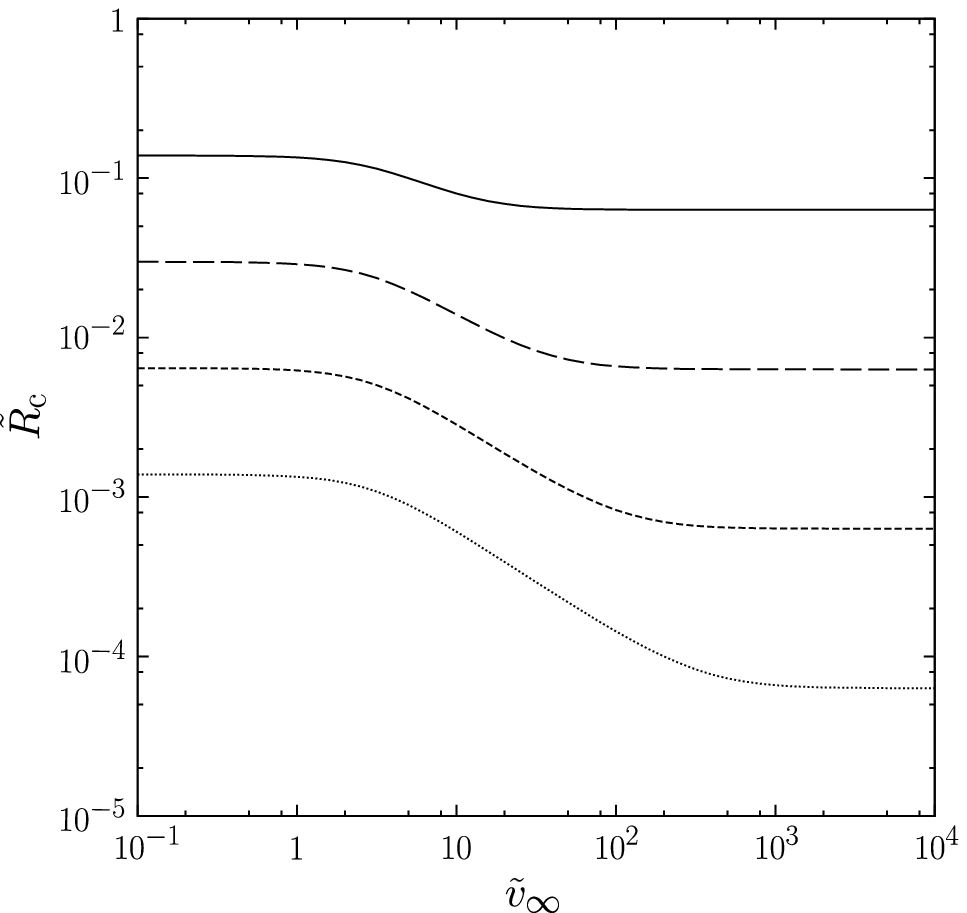}
\caption{Enhanced radius $\tilde{R}_{\rm c}$ as a function of
$\tilde{v}_\infty$ when $\alpha=3$ and $f_{\rm c}=2$.  Four lines
correspond to the cases with $\xi = 10^{-3}, 10^{-5}, 10^{-7}, 10^{-9}$
from top to bottom.  Core radius in this normalization is typically
about 0.005 for 1AU and 0.001 for 5AU.
\label{fig_R_c}}
\end{figure}

The total specific energy of a particle on the Hill coordinate system is
given by
\begin{eqnarray}
\tilde{E}
&=& \tilde{\Phi} + \frac{1}{2}\tilde{v}^2 \nonumber \\
&=& \frac{1}{2}(\tilde{e}^2+\tilde{i}^2)
  - \frac{3}{8}\tilde{b}^2
  + \frac{9}{2},
\label{tilde_E}
\end{eqnarray}
where $\tilde{v}$ is the velocity of the particle, and $\tilde{E}$
corresponds to the Jacobi integral \citep{Nakazawa1989}.  Note that the
orbital elements $\tilde{e}$, $\tilde{i}$, and $\tilde{b}$ in the last
expression of Eq.(\ref{tilde_E}) are the initial unperturbed values,
thus the term for mutual gravity $(-3/\tilde{r})$ is neglected in this
expression.
Since the $\tilde{\Phi} = 0$ contour surface is closed, a planetesimal
cannot escape from the planet's Hill sphere, if the total energy
$\tilde{E}$ becomes negative due to gas drag \citep{Ohtsuki1993}.
Therefore, the condition for capture by gas drag is given
by\footnote{The corresponding condition given by \citet{II03}, namely
$\Delta \tilde{E} > (1/2)\tilde{v}_\infty^2 + 3$ in our notation, is
slightly different from Eq.~(\ref{Delta_E_inequality}), because they did
not consider the contribution of the tidal potential.}
\begin{equation}
\Delta \tilde{E}
 > \frac{1}{2}\tilde{v}_\infty^2 + \frac{9}{2}.
\label{Delta_E_inequality}
\end{equation}
The critical value of the minimum distance $\tilde{r}_{\rm min}$
obtained by equating both sides of Eq.(\ref{Delta_E_inequality}) is
denoted by $\tilde{R}_{\rm c}$, which can be regarded as the {\it
enhanced radius} of a planet inside which incoming objects get
effectively captured by the gas drag \citep{II03}.  Thus, using
Eqs.(\ref{Delta-E}) and (\ref{Delta_E_inequality}), the enhanced radius
is obtained as the solution of
\begin{equation}
2f_{\rm c} \xi
\frac{\tilde{v}_\infty^2+6/\tilde{R}_{\rm c}}
     {\tilde{v}_\infty^2+9}
\tilde{R}_{\rm c}^{1-\alpha}
- 1 = 0,
\label{R_c_equation}
\end{equation}
and is shown in Fig. \ref{fig_R_c} as a function of $\tilde{v}_\infty$.
Depending on $\tilde{v}_\infty$, $\tilde{R}_{\rm c}$ can be divided into
three regimes and approximated by
\begin{equation}
\tilde{R}_{\rm c}
 \simeq
   \begin{cases}
    (4f_{\rm c}\xi/3)^{1/\alpha}
      & \mbox{for $\tilde{v}_\infty^2 < 9$}, \\
    (12f_{\rm c}\xi \tilde{v}_\infty^{-2})^{1/\alpha}
      & \mbox{for $9 < \tilde{v}_\infty^2
                    < 6(2f_{\rm c}\xi)^{-1/(\alpha-1)}$}, \\
    (2f_{\rm c}\xi)^{1/(\alpha-1)}
      & \mbox{for $6(2f_{\rm c}\xi)^{-1/(\alpha-1)} < \tilde{v}_\infty^2$}.
   \end{cases}
\label{R_c_approx}
\end{equation}
In the low velocity regime, the effect of solar gravity is important.
In the intermediate velocity regime, the two-body approximation is valid
\citep[e.g.,][]{GL90} and $\tilde{v}_\infty$ is smaller than the escape
velocity at $\tilde{r}=\tilde{R}_{\rm c}$ as well.  In the high velocity
regime, $\tilde{v}_\infty$ is larger than the escape velocity at
$\tilde{r}=\tilde{R}_{\rm c}$.

Once we obtain the enhanced radius, the capture probability $p_{\rm
cap}$ can be obtained by replacing $\tilde{r}_{\rm p}$ in
Eq.~(\ref{integrand_of_pcol}) with $\tilde{R}_{\rm c}$, namely
\begin{equation}
p_{\rm cap}(\tilde{e},\tilde{i},\tilde{b})
 = \frac{2\tilde{R}_{\rm c}^2}{3\pi |\tilde{b}| \tilde{i}}
   \sqrt{\frac{\tilde{e}^2+\tilde{i}^2-(3/4)\tilde{b}^2}
              {\tilde{e}^2-\tilde{b}^2}}
   \left( 1 + \frac{6/\tilde{R}_{\rm c}}
                   {\tilde{e}^2+\tilde{i}^2-(3/4)\tilde{b}^2}
   \right).
\label{integrand_of_pcap}
\end{equation}
Then, the capture rate $P_{\rm cap}$ can be obtained by integrating
Eq.(\ref{integrand_of_pcap}) with respect to $\tilde{b}$, as in the case
of $P_{\rm col}$ (Eq.(\ref{Pcol})).  Note that the integration cannot be
done analytically because $\tilde{R}_{\rm c}$ is not a simple function
but a solution of Eq.~(\ref{R_c_equation}), and is a function of
$\tilde{b}$ through $\tilde{v}_\infty$ (Eq.(\ref{v_infty})).

\section{Numerical simulation}\label{sec_numerical_simulation}
\subsection{Numerical Method}
We integrate Eq.~(\ref{EOM}) for particles with various initial
orbital elements, using the eighth-order Runge-Kutta integrator.
For a given $(\tilde{e},\tilde{i})$, we calculate a phase volume for
accretion in the $(\tilde{b},\tau,\varpi)$ phase space by finding orbits
that lead to accretion.  For high velocity cases, the phase volume is
much smaller than the total volume, and we need to calculate a large
number of orbits to obtain $P_{\rm acc}$ with sufficient accuracy.  We
thus adopt a kind of adaptive mesh refinement method for $\tau$ and
$\varpi$ mesh; phase volume for accretion is narrowed down by several
steps of calculations from a coarser mesh with a larger target to a
finer mesh with a smaller target \citep{Ohtsuki1993}.
In our numerical code, the initial azimuthal distance is set to
$\tilde{y}_0 = 50$; we confirmed that this is large enough to obtain
accretion rates with sufficient accuracy.  We stop our orbital
integration when one of the following conditions is met:
(i) $\tilde{r} > \tilde{y}_0 + \tilde{b}_0 + 2\tilde{e}_0 + \tilde{i}_0$
(the subscript 0 denotes initial values),
(ii) $\tilde{r}<\tilde{r}_{\rm p}$,
(iii) $\tilde{E}<0$.
\footnote{ To be captured by the protoplanet's Hill sphere, one may
think that the third condition should be $\tilde{E}<0$ and $\tilde{r}<1$
as well, but we consider the range of initial orbital elements
corresponding to $\tilde{E}>0$ because planetesimals cannot enter the
Hill sphere if $\tilde{E}<0$ initially, and also we set the gas drag
term effective only within the Hill sphere, that is, where $\tilde{r}<1$
and $\tilde{\Phi}<0$, and $\tilde{E}$ decreases only by gas drag, thus
the condition $\tilde{r}<1$ is automatically met when $\tilde{E}$
decreases to a negative value.}
Hereafter, we set $\alpha=3$ unless otherwise stated.

Gas velocity is assumed to be zero on the rotating coordinate system,
i.e., $\Delta \tilde{\BF{u}} = \tilde{\BF{v}}$ where $\tilde{\BF{v}}$ is
the velocity vector of particles, since the gas velocity at a distance
$\tilde{r}$ from the planet is much smaller than the Keplerian velocity
around the planet ($=\sqrt{3/\tilde{r}}$) at the location, which is the
typical velocity of a planetesimal moving around the planet under the
strong effect of the planet's gravity.  As we consider atmospheres in
hydrostatic equilibrium before runaway gas accretion, gas velocity is
much smaller than the sound speed.  Also, the sound speed in an
atmosphere well inside of the Bondi radius is always smaller than the
Keplerian velocity.  Thus, the neglect of gas velocity is consistent
with the assumption of atmospheres in hydrostatic equilibrium.
We do not consider ablation of incoming particles when they are passing
through an atmosphere.  Although ablation may be important for
atmospheres of massive protoplanets \citep[$\geq
10M_\oplus$;][]{Benvenuto2008}, its effect can be neglected for
atmospheres before runaway gas accretion considered in the present work
\citep{II03}.

\subsection{Effects of gas drag}\label{sec_gas_drag}
Before discussing numerical results of accretion rates in detail, first
we show some examples of orbital calculations, which help us understand
the basic dynamical behavior of planetesimals affected by gas drag
in atmospheres.

\begin{figure}
\epsscale{0.95}
\plotone{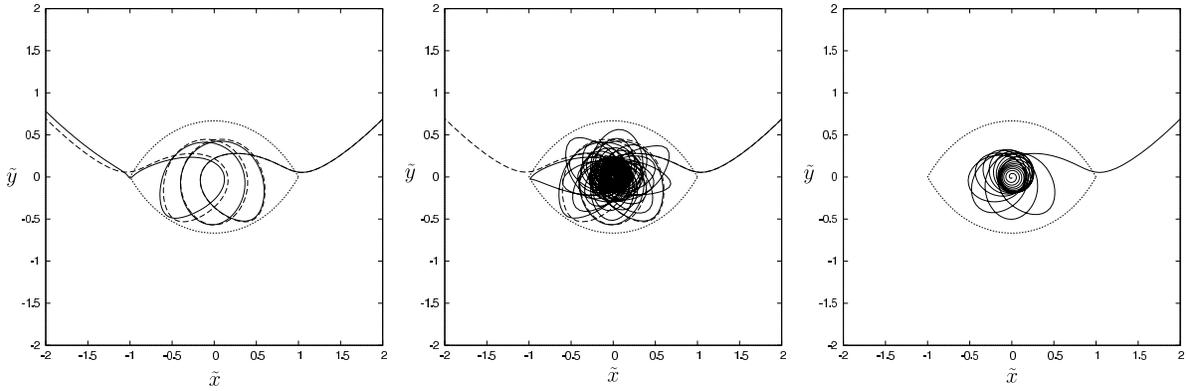}
\caption{Example orbits to show the effect of gas drag when $\tilde{b} =
4.855$, $\tilde{e} = 3.0$, $\tilde{i} = 0$, $\tau=0.046$.  Solid lines
show the orbits with gas drag when $\xi = 1.66\e{-5}$, $1.67\e{-5}$,
$2.00\e{-4}$, respectively from left to right.  Dashed lines in the left
and middle panels show the orbit without gas drag.  Dotted line
represents the Hill sphere.
\label{fig_orbits}}
\end{figure}

Figure \ref{fig_orbits} shows examples of coplanar orbits under gas drag
of atmospheres, with three different values of the gas drag parameter
$\xi$.  In the case with $\xi=1.66\e{-5}$ (left panel), the orbit is
slightly affected by gas drag, but the energy dissipation was not
significant and the object escapes from the Hill sphere.  When
$\xi=1.67\e{-5}$ (middle panel), the particle loses enough energy by gas
drag to be captured within the Hill sphere.  When $\xi=2\e{-4}$, the
particle loses a large amount of energy at the first encounter, and
spirals toward the planet quickly.

\begin{figure}
\epsscale{0.7}
\plotone{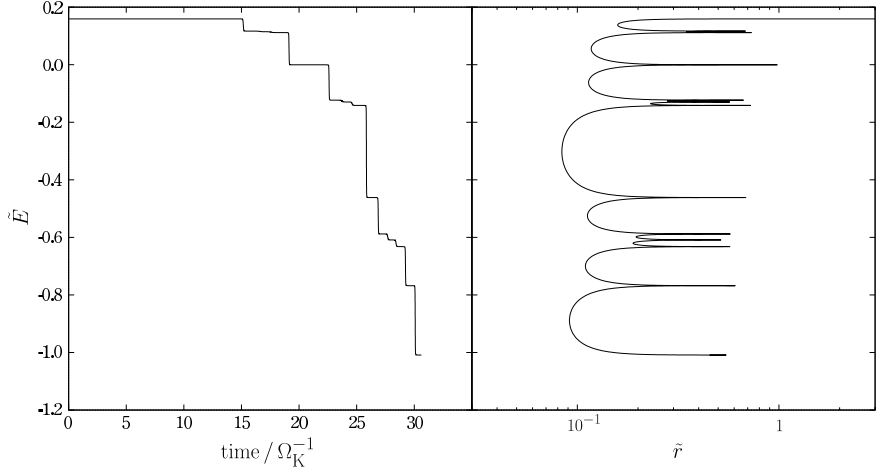}
%
\caption{Change of the Jacobi energy of a particle due to gas drag, as a
function of time (left panel) and the distance from the origin (right
panel), for the orbit shown in the middle panel of Fig.\ref{fig_orbits}
($\xi=1.67\e{-5}$).
\label{fig_Delta-J_vs_t-and-r}}
\end{figure}

Next, we examine the change of energy of a particle.  Figure
\ref{fig_Delta-J_vs_t-and-r} shows the change of $\tilde{E}$ for the
captured orbit described in the middle panel of Fig.~\ref{fig_orbits}.
The left and right panels show the plots of $\tilde{E}$ as a function of
time and the distance from the planet, respectively.  We find from the
right panel that the energy reduces greatly when the particle passes
through the dense part of the atmosphere near the planet's surface
(i.e., $\tilde{r} \leq 0.1$ in the case shown in
Fig. \ref{fig_Delta-J_vs_t-and-r}), whereas it is almost constant when
it is far from the planet.
Also, the left panel shows that significant energy decrease occurs
instantaneously, because most of energy dissipation occurs near
pericenters, where particles move with the fastest velocity in an orbit.
Note that, for the purpose of demonstration, the condition (iii) for
stopping integration described above was not applied to the orbital
integrations shown in Figs.~\ref{fig_orbits} and
\ref{fig_Delta-J_vs_t-and-r}.

\begin{figure}
\epsscale{0.4}
\plotone{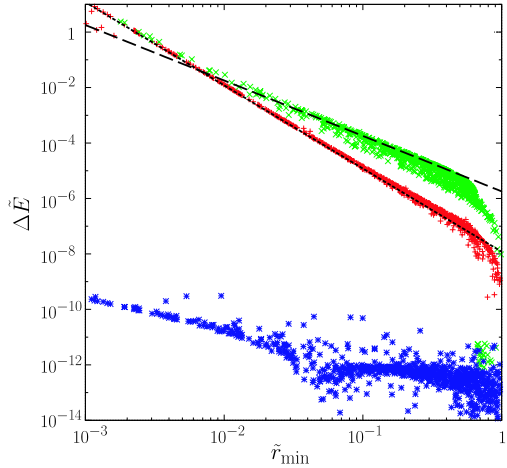}
\caption{Total change of the energy $\tilde{E}$ due to gas drag through
an encounter with the planet for each orbit, as a function of the
minimum distance to the planet center in the case with $\alpha=3.113$.
Red pluses and green crosses show the low-velocity ($\tilde{e}=3$) and
high-velocity ($\tilde{e}=30$) cases, respectively.  For each value of
$\tilde{e}$, numerical results for $\xi = 1\e{-9}$ with a wide range of
initial phase angle $\tau$ and impact parameter $\tilde{b}$ in the
two-dimensional case ($\tilde{i}=0$) are shown.  Blue asterisks show the
case without gas drag, that is, the error of the numerical integration.
Long-dashed line shows the approximate analytic results
(Eq.(\ref{Delta-E})) for the high-velocity regime (i.e., the term
proportional to $\tilde{r}_{\rm min}^{1-\alpha}$) and short-dashed line
shows that for the low-velocity regime (the term proportional to
$\tilde{r}_{\rm min}^{-\alpha}$).  We set $f_{\rm c}=2$.
\label{fig_Delta-J}}
\end{figure}

Figure \ref{fig_Delta-J} shows the total change of energy $\tilde{E}$
due to gas drag through an encounter with the planet until one of the
truncation conditions for orbital integration (i)--(iii) is met, as a
function of the minimum approach distance from the planet's center.  Two
groups of points with $\Delta \tilde{E} \gtrsim 1\e{-9}$ (red pluses and
green crosses) show numerical results for the low- and high-velocity
cases, respectively.
We can see that the energy dissipation is well described by the analytic
estimates given by Eq.~(\ref{Delta-E}), which are shown by the two lines
(we set $f_{\rm c} = 2$ in Eq.(\ref{Delta-E})).  For the low velocity
case, $\Delta \tilde{E} \propto \tilde{r}_{\rm min}^{-\alpha}$, which
corresponds to the second term of Eq.~(\ref{Delta-E}) and means that the
particles are well accelerated by the planet gravity at $\tilde{r} \sim
\tilde{r}_{\rm min}$.  For the high velocity case, $\Delta \tilde{E}
\propto \tilde{r}_{\rm min}^{1-\alpha}$, which corresponds to the first
term of Eq.~(\ref{Delta-E}) and means that the particle velocity around
$\tilde{r} \sim \tilde{r}_{\rm min}$ is not much larger than
$\tilde{v}_\infty$.
We also plot the changes of $\tilde{E}$ for calculations without gas
drag, which corresponds to the numerical error in our numerical
integration.  We find that the numerical error is much smaller than the
calculated energy changes due to gas drag, even in the case when gas
drag is very weak ($\xi=1\e{-9}$).

\begin{figure}
\epsscale{0.9}
\plottwo{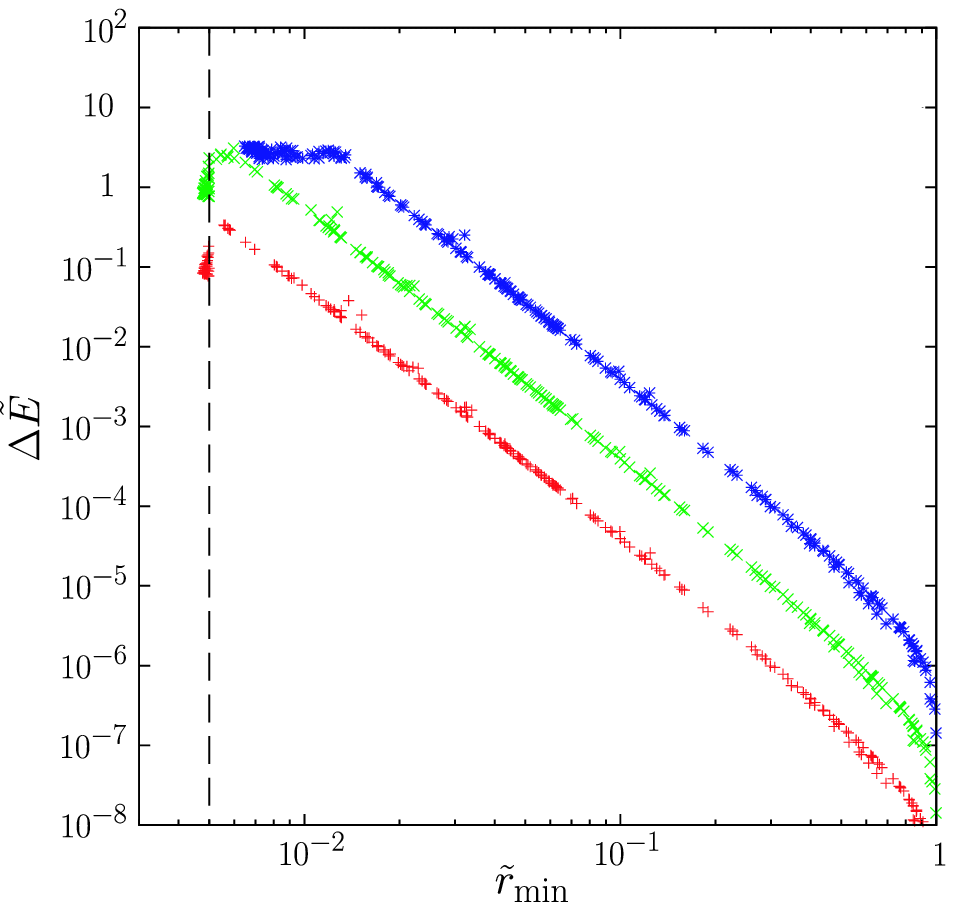}{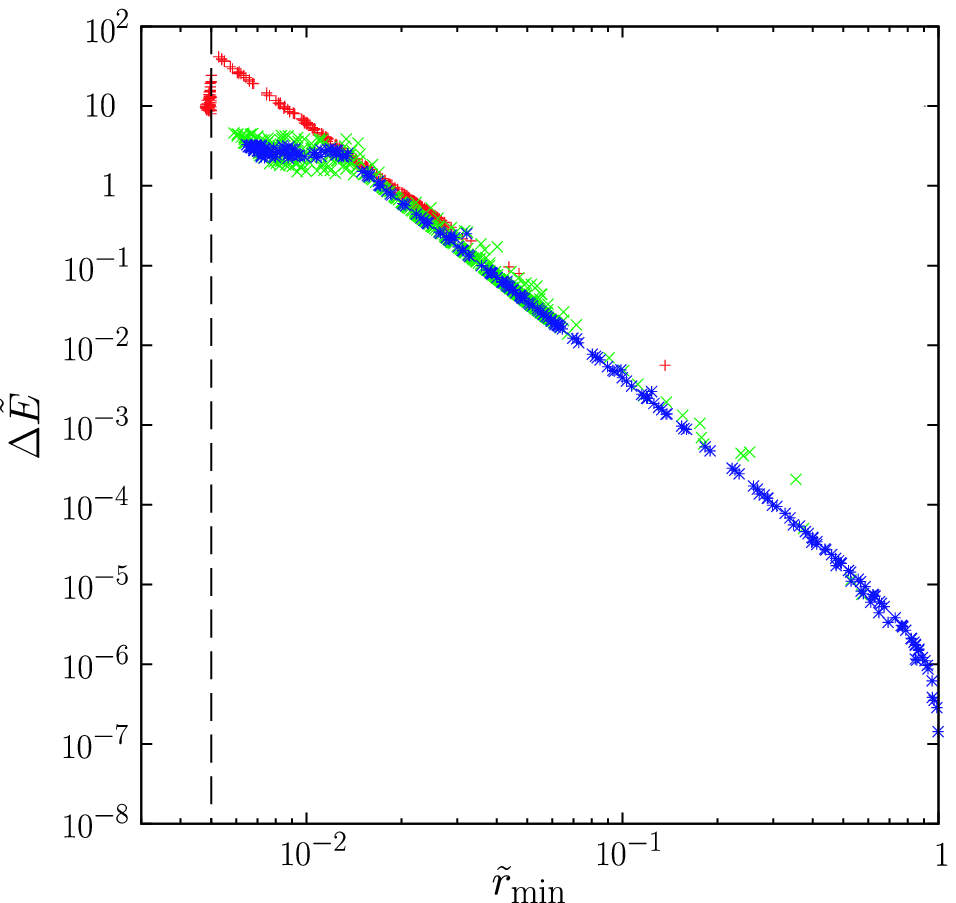}
\caption{Total change of the energy $\tilde{E}$ due to gas drag through
an encounter with the planet for each orbit, as a function of the
minimum distance in the case with $\alpha=3.113$.  Left panel shows the
dependence on $\xi$ when $(\tilde{e},\tilde{i}) = (0.1, 0.05)$: red
pluses, green crosses, and blue asterisks are the cases with
$\xi=2.56\e{-9}$ (weak), $2.56\e{-8}$ (intermediate), and $2.56\e{-7}$
(strong), respectively.  Right panel shows the dependence on initial
orbital elements when $\xi=2.56\e{-7}$: red pluses, green crosses, and
blue asterisks are the case with $(\tilde{e},\tilde{i}) = (10,5)$,
$(1,0.5)$, and $(0.1,0.05)$, respectively.  The vertical dashed line
shows the planet's radius $\tilde{r}_{\rm p}=5\e{-3}$.
\label{fig_Delta-J_xi-e-depend}}
\end{figure}

In Fig. \ref{fig_Delta-J_xi-e-depend}, we show the total change of
$\tilde{E}$ in the same way as Fig.~\ref{fig_Delta-J} in the case with
$\tilde{r}_{\rm p} = 0.005$ to examine the dependence on $\xi$ and
$\tilde{e}$.  Note that there are cases when the second or a later close
approach reduces energy by an amount comparable to or larger than that
due to the first one even when planetesimals are not captured within the
Hill sphere yet, but such events are rare unless planetesimals are
captured by the first close approach.  Thus, the values of $\Delta
\tilde{E}$ shown in Fig.~\ref{fig_Delta-J_xi-e-depend} practically
represent the energy change due to the first close approach to the
planet, which can be used as an indicator whether planetesimals are
captured or not from the point of view of energetics.
The left panel shows the dependence on $\xi$ for a low-velocity case.
In this cases, we find $\Delta \tilde{E} \propto \tilde{r}_{\rm
min}^{-\alpha}$ for most points, as explained above.  In the weak gas
drag case, planetesimals reduce their energy due to single close
encounter and escape from the Hill sphere.
In the strong gas drag case, points shift upward in accordance with
stronger energy dissipation.  The points in this case fall on a
horizontal band with $\Delta \tilde{E} \sim$ 2--3 at $\tilde{r}_{\rm
min} \lesssim 0.015$; in this regime, particles are captured by gas drag
rather than direct collision onto the planet, i.e., orbital integration
is stopped by the condition $\tilde{E}<0$.  Thus, values of $\Delta
\tilde{E}$ in this horizontal band in the plot show the amount of energy
that needs to be dissipated for capture, and the upper end of this
horizontal band ($\tilde{r}_{\rm min} \sim 0.015$) corresponds to
$\tilde{R}_{\rm c}$ (see Eq.~(\ref{R_c_approx})).
When gas drag is weaker, energy dissipation becomes smaller and
$\tilde{R}_{\rm c}$ becomes smaller accordingly, and eventually the
trend $\Delta \tilde{E} \propto \tilde{r}_{\rm min}^{-\alpha}$ does not
reach the energy needed to be captured ($\Delta \tilde{E}\sim$ 2--3 in
this case) before it reaches the surface of the planet ($\tilde{r} =
\tilde{r}_{\rm p}$); in this case, particles are accreted by direct
collision.  Note that the points with $\tilde{r}_{\rm min} \lesssim
0.005$ represent cases where orbital integrations were stopped by the
condition of direct collision to the planet.
The right panel shows the dependence on random velocity (i.e.,
$\tilde{e}$ and $\tilde{i}$).  The low velocity case is identical to the
strong drag case in the left panel.  With increasing random velocity,
the energy needed to be captured increases because initial kinetic
energy is larger, and $\tilde{R}_{\rm c}$ decreases accordingly.  In the
high velocity case, the numerical results fall on a line with a nearly
constant slope which extends all the way to $\tilde{r}_{\rm min} =
\tilde{r}_{\rm p}$, which means that gas drag is not strong for
particles with such a high random velocity and they are accreted only by
direct collision with the planet's solid surface.

\subsection{Accretion rates for the case with $\tilde{e}=2\tilde{i}$}
\begin{figure}
\epsscale{0.7}
\plotone{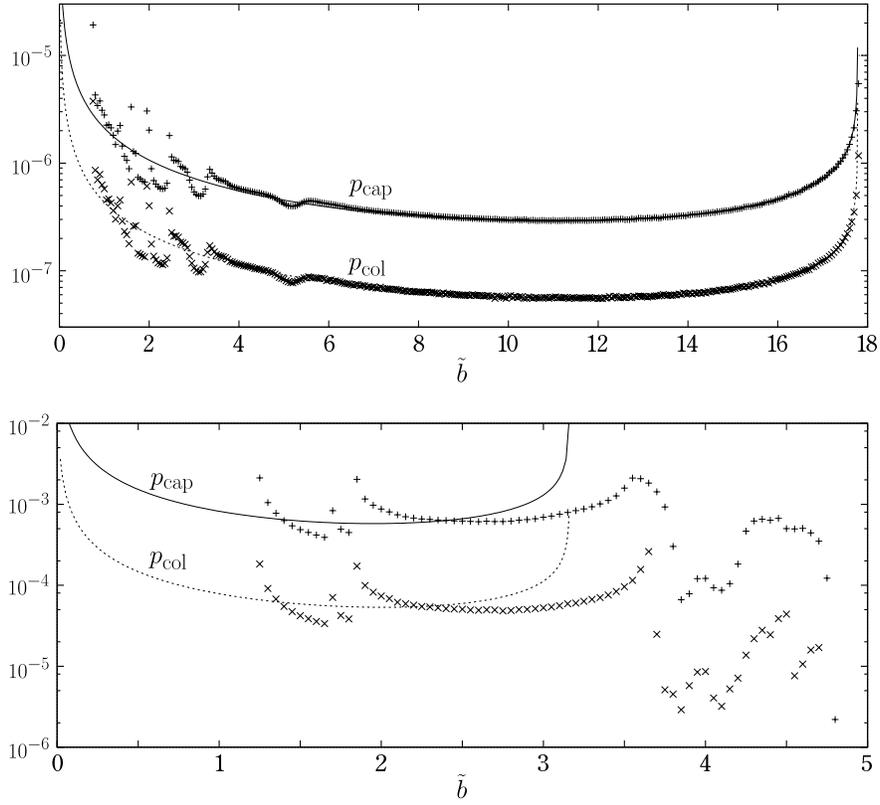}
\caption{ Differential non-dimensional accretion rate $p_{\rm cap}$ and
collision rate $p_{\rm col}$ as a function of $\tilde{b}$.  Top and
bottom panels shows $(\tilde{e},\tilde{i}) = (17.8,8.9)$ and
$(3.16,1.58)$, respectively.  Solid and dotted lines show analytic
solutions of $p_{\rm cap}$ and $p_{\rm col}$, respectively.  Pluses and
crosses show numerical solutions of $p_{\rm cap}$ and $p_{\rm col}$,
respectively.
\label{fig_dp-cap}}
\end{figure}

Next, we examine accretion rates in the case of $\tilde{e}=2\tilde{i}$.
Figure \ref{fig_dp-cap} shows the plots of $p_{\rm col}
(\tilde{e},\tilde{i},\tilde{b})$ and $p_{\rm cap}
(\tilde{e},\tilde{i},\tilde{b})$ as a function of $\tilde{b}$, for two
different sets of $(\tilde{e},\tilde{i})$.
Crosses and pluses represent numerical results for $p_{\rm col}$
(Eq.(\ref{integrand_of_pcol})) and $p_{\rm cap}$
(Eq.(\ref{integrand_of_pcap})), respectively, and the dotted and solid
lines show the analytic results.
Note that we empirically set $f_{\rm c}=1.83$ for the analytic results
of $p_{\rm cap}$; changing $f_{\rm c}$ only shifts the lines in the
vertical direction and does not change the functional form.
In the high $(\tilde{e},\tilde{i})$ case, we find that the numerical
results are in good agreement with the analytic solutions over a wide
range of $\tilde{b}$, except for the small-$\tilde{b}$ region where some
small deviations are observed.
In the low $(\tilde{e},\tilde{i})$ case, numerical results deviate from
the analytic solution significantly.  This is because $\tilde{v}_\infty$
is not large enough to neglect the three-body effect in this case.

\begin{figure}
\epsscale{0.9}
\plottwo{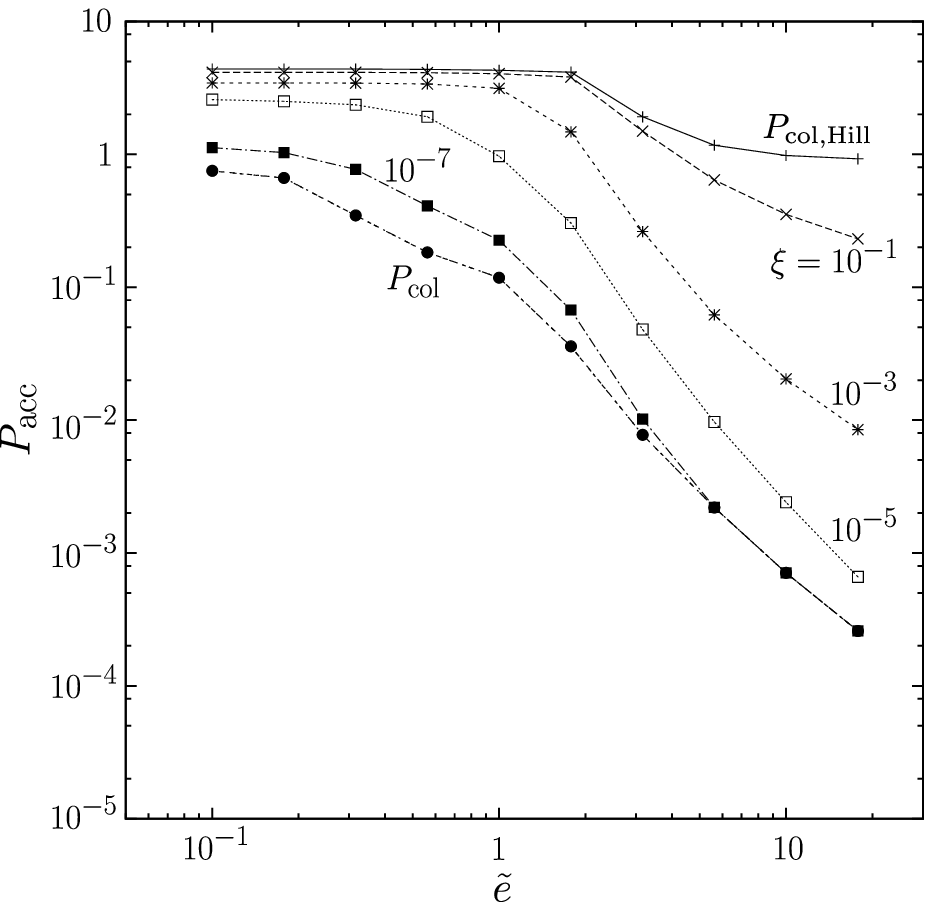}{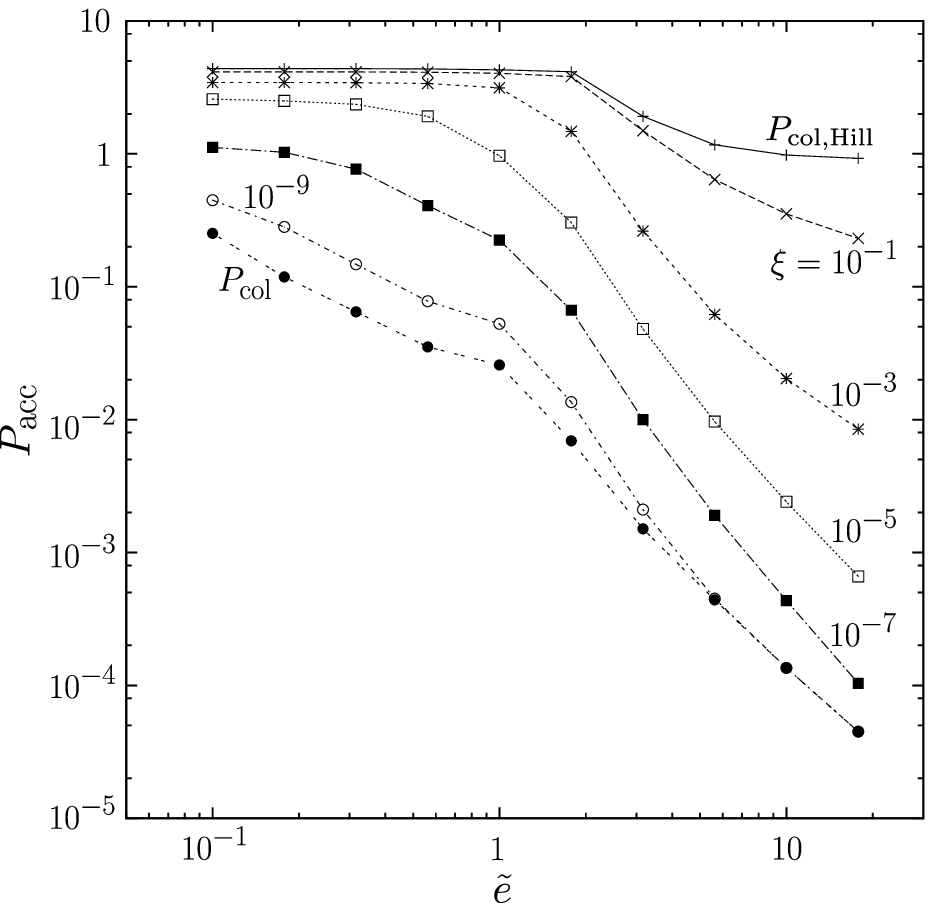}
\caption{ Normalized accretion rate $P_{\rm acc}$ as a function of
$\tilde{e}$ when $\tilde{e}=2\tilde{i}$ for several values of $\xi$.
Left panel shows the case with $\tilde{r}_{\rm p}=0.005$ and right panel
shows when $\tilde{r}_{\rm p}=0.001$.  The uppermost line in both panels
shows normalized collision probability onto the lemon-shaped surface of
the Hill sphere.  The bottom line in both panels shows $P_{\rm col}$.
\label{fig_Pacc_xi-13579}}
\end{figure}

Figure \ref{fig_Pacc_xi-13579} shows numerical results of $P_{\rm acc}$
for several values of $\xi$, along with $P_{\rm col}$ (bottom line).  We
also plot numerical results for $P_{\rm col,Hill}$, which is the
collision rate onto the Hill sphere (top line).
All these curves can be roughly divided into two parts: a horizontal
part in the low-velocity regime and a part with a negative slope in the
high-velocity regime.
In the lower eccentricity regime, previous studies showed that $P_{\rm
col}$ is independent of $\tilde{v}_\infty$ and depends on
$\tilde{r}_{\rm p}$ as \citep{IN89,Inaba2001}
\begin{equation}
P_{\rm col,low}
 = 11.3 \sqrt{\tilde{r}_{\rm p}}.
\label{Pcol_low}
\end{equation}
This corresponds to the horizontal part of the curve for $P_{\rm col}$
in Fig.~\ref{fig_Pacc_xi-13579}, even though the horizontal part can be
seen only in the lowest $\tilde{e}$ regime displayed in the left figure.
In the high eccentricity regime, $P_{\rm col}$ can be well described by
Eq.~(\ref{Pcol}) and is thus proportional to $\tilde{e}^{-2}$; this
corresponds to the negative-slope part of the above curve.

\begin{figure}
\epsscale{0.4}
\plotone{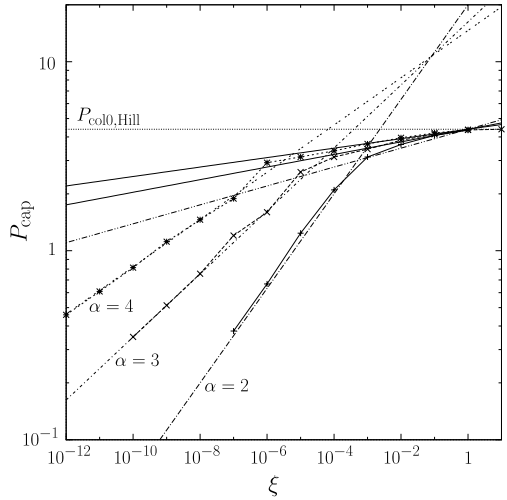}
\caption{ Normalized capture rate as a function of $\xi$ when $\tilde{e}
 = \tilde{i} = 0$.  Numerical results for three values of $\alpha$ are
 shown by symbols.  Six fit lines for these results described by
 Eqs.~(\ref{Pcap0_low_xi}) and (\ref{Pcap0_med_xi}) are also shown.
\label{fig_Pcap_vs_xi_ei0}}
\end{figure}

In the case for planets with atmospheres, this general trend still
holds, but changes in some parts.
In the lower eccentricity regime, $P_{\rm cap}$ is expected to be
proportional to $\xi^{1/2\alpha}$, because $P_{\rm cap} \propto
\tilde{R}_{\rm c}^{1/2}$ (see Eq.~(\ref{Pcol_low})) and $\tilde{R}_{\rm
c} \simeq (4f_{\rm c}\xi/3)^{1/\alpha}$ (see Eq.~(\ref{R_c_approx})).
On the other hand, Fig.~\ref{fig_Pacc_xi-13579} shows a rather weak
dependence on $\xi$ in the low-velocity regime, especially when $\xi
\geq 10^{-5}$.  In order to examine the $\xi$-dependence in the
low-velocity regime in detail, we calculated capture rates as a function
of $\xi$ for $\tilde{e} = \tilde{i} = 0$
(Fig.~\ref{fig_Pcap_vs_xi_ei0}).  In the low $\xi$ regime ($\xi \lesssim
10^{-5}$ when $\alpha=3$), $P_{\rm cap}$ is indeed $\propto
\xi^{1/2\alpha}$.
In the high $\xi$ regime ($\xi>1$), on the other hand, $P_{\rm cap}$ is
constant and is equal to the collision rate onto the Hill sphere,
because particles entering the Hill sphere immediately get captured due to
the strong gas drag.
In between the above two extreme regimes of $\xi$ in
Fig.~\ref{fig_Pcap_vs_xi_ei0}, there is a part where $P_{\rm cap}$ is
well approximated by $\xi^{1/10\alpha}$ at $10^{-5} \lesssim \xi
\lesssim 1$ when $\alpha=3$.  We thus obtain the following empirical
formula for capture rates in the low velocity regime
\begin{equation}
P_{\rm cap0}
 = \min(P_{\rm cap0,low\xi}, P_{\rm cap0,med\xi}, P_{\rm col0,Hill}),
\label{Pcap0}
\end{equation}
where $P_{\rm cap0,low\xi}$ is the capture rate in the low $\xi$ regime,
$P_{\rm cap0,med\xi}$ is the one for the intermediate $\xi$ regime, and
$P_{\rm col0,Hill}$ is the collision rate onto the Hill sphere,
respectively, and these variables are defined in the case with
$\tilde{e}=\tilde{i}=0$, given by
\begin{equation}
P_{\rm cap0,low\xi}
 = 10.7
   \kakkoi{\xi}{0.082}{1/2\alpha},
\label{Pcap0_low_xi}
\end{equation}
\begin{equation}
P_{\rm cap0,med\xi}
 = P_{\rm col,Hill} \xi^{1/10\alpha},
\label{Pcap0_med_xi}
\end{equation}
\begin{equation}
P_{\rm col0,Hill} = 4.392.
\label{Pcol0_Hill}
\end{equation}
As we can see in Fig.~\ref{fig_Pcap_vs_xi_ei0}, the expressions
(\ref{Pcap0})--(\ref{Pcol0_Hill}) reproduce our numerical results very
well, even when $\alpha = 2$ and 4.

In the high-velocity regime, we can see from
Fig.~\ref{fig_Pacc_xi-13579} that, when $\tilde{e} \sim 10$ and $\xi$ is
small, the $\tilde{e}$-dependence of $P_{\rm cap}$ is stronger than that
of $P_{\rm col}$ ($\propto \tilde{e}^{-2}$), because $\tilde{R}_{\rm c}$
decreases with increasing random velocity in this regime (see
Eq.~(\ref{R_c_approx})).  Thus, with increasing $\tilde{e}$, $P_{\rm
acc}$ for small $\xi$ approaches $P_{\rm col}$ at a certain $\tilde{e}$
($\tilde{e}\sim 5$ when $\xi=10^{-7}$ and $\tilde{r}_{\rm p}=0.005$, for
example).
In \S \ref{sec_empirical_formula}, we will derive an analytic expression
of the capture rate in this regime.  The expression shows that the
capture rate is expected to be proportional to $\tilde{e}^{-8/3}$ when
$\alpha=3$ and $\tilde{e}=2\tilde{i}$, which agrees well with the above
numerical results.

\subsection{Accretion rates averaged over the distribution of
  eccentricities and inclinations}
Planetesimals in protoplanetary disks have a distribution of
eccentricities and inclinations that is well described by the Rayleigh
distribution function \citep{IM92}, thus we next show $P_{\rm acc}$ for
incoming planetesimals that have the Rayleigh distribution in
eccentricity and inclination.

Normalized accretion rate for planetesimals that have a Rayleigh
distribution in $\tilde{e}$ and $\tilde{i}$ with the root mean square
values $\tilde{e}^\ast$ and $\tilde{i}^\ast$ is given by
\begin{equation}
\ave{P_{\rm acc}}
 \equiv \iint P_{\rm acc}
              f_{\rm R}(\tilde{e},\tilde{i}, \tilde{e}^\ast, \tilde{i}^\ast)
 d\tilde{e} d\tilde{i},
\label{Pacc_ave_define}
\end{equation}
where $f_{\rm R}$ is the Rayleigh distribution function given by
\begin{equation}
f_{\rm R}(\tilde{e},\tilde{i},\tilde{e}^\ast,\tilde{i}^\ast)
 = \frac{4\tilde{e}\tilde{i}}
        {(\tilde{e}^\ast)^2 (\tilde{i}^\ast)^2}
   \exp\left( -\frac{\tilde{e}^2}{(\tilde{e}^\ast)^2}
              -\frac{\tilde{i}^2}{(\tilde{i}^\ast)^2}
       \right).
\end{equation}
In order to obtain $\ave{P_{\rm acc}}$ from Eq.~(\ref{Pacc_ave_define})
for certain regimes of $\tilde{e}^\ast$ and $\tilde{i}^\ast$ ($10^{-1}
\leq \tilde{e}^\ast \leq 10^{5/4}$ and $10^{-1} \leq \tilde{i}^\ast \leq
10^{5/4}$ in this paper), we need values of $P_{\rm acc}$ for a wide
range of $\tilde{e}$ and $\tilde{i}$ including the region outside of the
regime of $\tilde{e}^\ast$ and $\tilde{i}^\ast$ \citep{Ohtsuki1999}.  To
do that, we follow the method described in \citet{Ohtsuki1999} with a
slight modification.

We first obtain $P_{\rm acc}(\tilde{e},\tilde{i})$ by direct orbital
integration described in the previous sections at the $10\times 10$ mesh
points in the square region $10^{-1} \leq \tilde{e} \leq 10^{5/4}$ and
$10^{-1} \leq \tilde{i} \leq 10^{5/4}$ with even interval in the
logarithmic space.  We also perform orbital integration at supplementary
coarse grid points in the low inclination regime, i.e., $3\times 2$ mesh
points with $\tilde{e} = 10^{-1}, 10^0, 10^1$ and $\tilde{i} = 10^{-3},
10^{-2}$.
Values outside of the region are obtained by extrapolation from the
calculated region and by the analytic estimation for the high-velocity
regime (see \S \ref{sec_empirical_formula}).  We confirmed that
$\ave{P_{\rm acc}}$ for $10^{-1} \leq \tilde{e}^\ast \leq 10^{5/4}$ does
not depend sensitively on the $P_{\rm acc}$ values for the outside of
the square region $10^{-1} \leq \tilde{e} \leq 10^{5/4}$ and $10^{-1}
\leq \tilde{i} \leq 10^{5/4}$.

\begin{figure}
\epsscale{0.9}
\plottwo{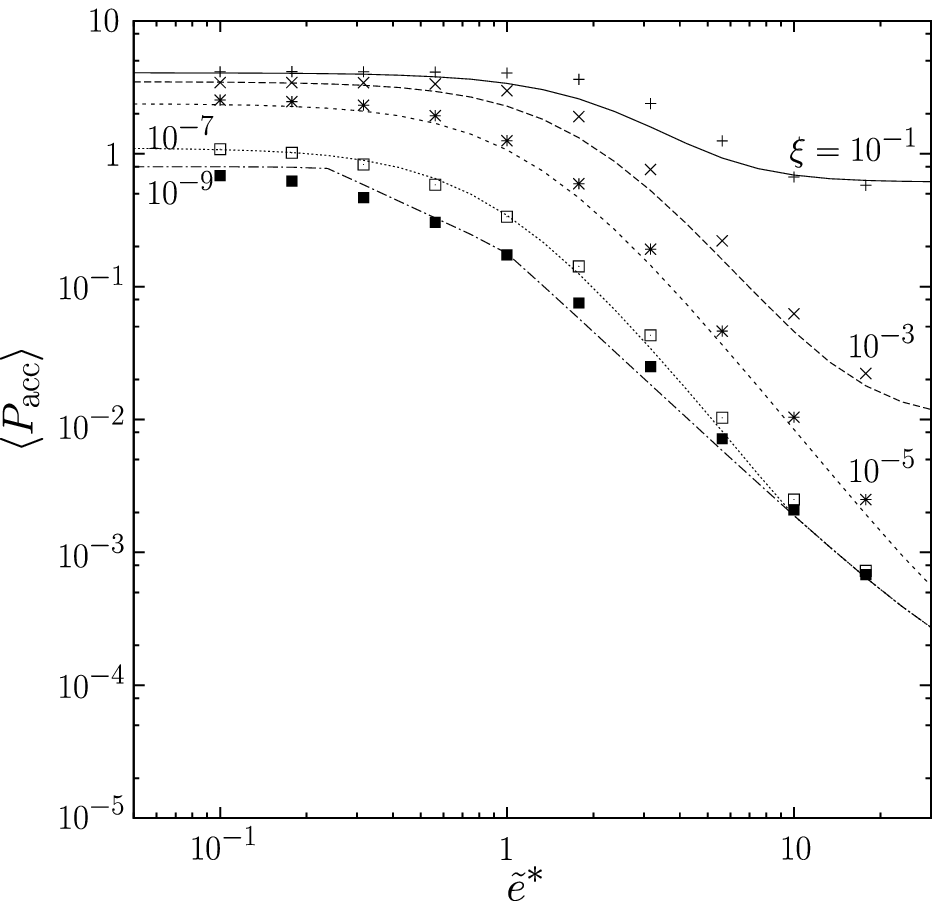}{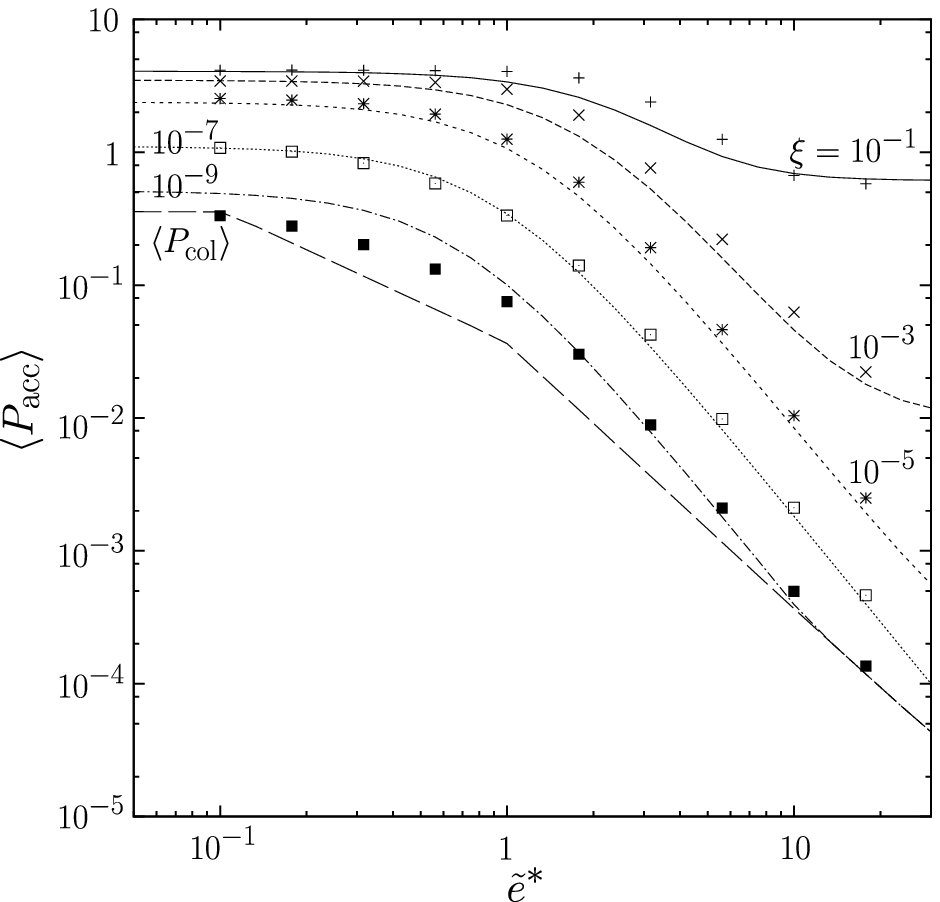}
\caption{ Normalized accretion rates averaged over the Rayleigh
distribution for $\tilde{r}_{\rm p}=0.005$ (left panel) and
$\tilde{r}_{\rm p}=0.001$ (right panel), as a function of
$\tilde{e}^\ast (= 2\tilde{i}^\ast)$.
Marks show simulation results for $\xi = 10^{-1}$, $10^{-3}$, $10^{-5}$,
$10^{-7}$, and $10^{-9}$, and lines show the semi-analytical formula
given by Eq.~(\ref{Pacc_ave}).  The Rayleigh distribution average of
collision rate (Eq.~(\ref{Pcol_ave})) is also shown by the long dashed
line for comparison in the case of $\tilde{r}_{\rm p} = 0.001$ (left
panel).  Note that, in the case of $r_{\rm p} = 0.005$ and
$\xi=10^{-9}$, there is no effect of the atmosphere on $\ave{P_{\rm
acc}}$, i.e., $\ave{P_{\rm acc}} = \ave{P_{\rm col}}$
\label{fig_Pacc_ave}}
\end{figure}

Symbols in Fig.~\ref{fig_Pacc_ave} shows numerical results for
$\ave{P_{\rm acc}}$ as a function of $\tilde{e}^\ast$ when
$\tilde{e}^\ast = 2\tilde{i}^\ast$.
While this figure looks similar to Fig.~\ref{fig_Pacc_xi-13579}, the
results shown in Fig.~\ref{fig_Pacc_ave} change more smoothly with
increasing random velocity than those of Fig.~\ref{fig_Pacc_xi-13579},
and the absolute values in Fig.~\ref{fig_Pacc_ave} tend to be slightly
higher than those of Fig.~\ref{fig_Pacc_xi-13579} in the high-velocity
regime.  These two differences come from the average over the Rayleigh
distribution, and similar results were obtained for the Rayleigh
distribution average of collision rates \citep{GL92}.

\section{A semi-analytic expression for accretion rates}
\label{sec_empirical_formula}
Next, we derive a semi-analytical expression for $\ave{P_{\rm acc}}$
based on the numerical and analytical calculations for accretion rates
described above.  As shown in Figs.~\ref{fig_Pacc_xi-13579} and
\ref{fig_Pacc_ave}, with decreasing strength of gas drag (i.e., $\xi$),
accretion rates decrease and approach the collision rates onto the solid
surface of the planet without an atmosphere.  Thus $\ave{P_{\rm acc}}$
can be approximated by the larger of $\ave{P_{\rm col}}$ and
$\ave{P_{\rm cap}}$:
\footnote{Note that, since the integrands in the definition of $P_{\rm
col}$ and $P_{\rm cap}$ (see Eq.(\ref{Pacc})) depend on $\tilde{b}$, the
magnitude relation between the two also depends on $\tilde{b}$, so a
more precise procedure would be to take $ p_{\rm acc} = \max (p_{\rm
col},p_{\rm cap})$ first and then integrate it with respect to
$\tilde{b}$, $\tilde{e}$, and $\tilde{i}$.  However, for the derivation
of an approximate formula for $\ave{P_{\rm acc}}$, the simple procedure
described here is sufficient.}
\begin{equation}
\ave{P_{\rm acc}}(\tilde{r}_{\rm p}, \xi, \alpha,
                  \tilde{e}^\ast, \tilde{i}^\ast )
 = \max (\ave{P_{\rm col}}(\tilde{r}_{\rm p},
              \tilde{e}^\ast, \tilde{i}^\ast),
         \ave{P_{\rm cap}}(\xi,\alpha, \tilde{e}^\ast, \tilde{i}^\ast)).
\label{Pacc_ave}
\end{equation}

Collision rate averaged over the Rayleigh distribution was studied in
detail by previous work \citep[e.g.,][]{GL92}, and \citet{Inaba2001}
derived the following semi-analytic expression for $\ave{P_{\rm col}}$:
\begin{equation}
\ave{P_{\rm col}}
 = \min \left( \ave{P_{\rm col}}_{\rm med},
               \left\{  \ave{P_{\rm col}}_{\rm low}^{-2}
                      + \ave{P_{\rm col}}_{\rm high}^{-2}
               \right\}^{-1/2}
        \right),
\label{Pcol_ave}
\end{equation}
where
\begin{equation}
\ave{P_{\rm col}}_{\rm low}
 = 11.3 \sqrt{\tilde{r}_{\rm p}},
\end{equation}
\begin{equation}
\ave{P_{\rm col}}_{\rm med}
 = \frac{\tilde{r}_{\rm p}^2}{4\pi \tilde{i}^\ast}
   \left( 17.3 + \frac{232}{\tilde{r}_{\rm p}}
   \right),
\label{Pcol_ave_med}
\end{equation}
\begin{equation}
\ave{P_{\rm col}}_{\rm high}
 = \frac{\tilde{r}_{\rm p}^2}{2\pi}
   \left( {\cal F}(I^\ast)
         +\frac{6}{\tilde{r}_{\rm p}}
          \frac{{\cal G}(I^\ast)}{(\tilde{e}^\ast)^2}
   \right).
\label{Pcol_ave_high}
\end{equation}
In the above, Eq.(\ref{Pcol_ave_high}) was obtained by averaging
Eq.~(\ref{Pcol}) over the Rayleigh distribution \citep{GL92}; $I^\ast
\equiv \tilde{i}^\ast/\tilde{e}^\ast$ and the function ${\cal
F}(I^\ast)$ and ${\cal G}(I^\ast)$ are given by
\begin{equation}
{\cal F}(I^\ast)
 \equiv 8\int_0^1
             d\lambda
             \frac{(I^\ast)^2 E[\sqrt{3(1-\lambda^2)}/2]}
                  {[(I^\ast)^2+(1-(I^\ast)^2)\lambda^2]^2},
\end{equation}
\begin{equation}
{\cal G}(I^\ast)
 \equiv 8\int_0^1
             d\lambda
             \frac{K[\sqrt{3(1-\lambda^2)}/2]}
                  {[(I^\ast)^2+(1-(I^\ast)^2)\lambda^2]}.
\end{equation}
When $I^\ast=1/2$, ${\cal F}(1/2) = 17.34$ and ${\cal G}(1/2) = 38.22$.

As for the capture rate averaged over the Rayleigh distribution, we
found that the following expression reproduces our numerical results
very well:
\begin{equation}
\ave{P_{\rm cap}}
 = \left(  \ave{P_{\rm cap}}_{\rm low }^{-\eta}
         + \ave{P_{\rm cap}}_{\rm high}^{-\eta}
   \right)^{-1/\eta},
\end{equation}
with $\eta = 2/3$.  The accretion rate for the low-velocity regime can
be simply written in terms of $P_{\rm cap0}$ (Eq.(\ref{Pcap0})) as
\begin{equation}
 \ave{P_{\rm cap}}_{\rm low} = P_{\rm cap0},
\end{equation}
because $P_{\rm cap0}$ is independent of $\tilde{e}$ and $\tilde{i}$.
We further divide $\ave{P_{\rm cap}}_{\rm high}$ into two functions as
\begin{equation}
\ave{P_{\rm cap}}_{\rm high}
 = \ave{P_{\rm cap}}_{\rm high1} + \ave{P_{\rm cap}}_{\rm high2}.
\end{equation}
We define that $\ave{P_{\rm cap}}_{\rm high1}$ is for a relatively
low-velocity regime in the sense that the relative velocity is low
enough for the collision cross-section to be enhanced by gravitational
focusing, whereas it is still high enough to hold the two-body
approximation.  On the other hand, $\ave{P_{\rm cap}}_{\rm high2}$ is
defined for a higher velocity regime where gravitational focusing is
negligible.  In the former regime, $P_{\rm cap,high1}$ can be obtained
by the integration of Eq.~(\ref{integrand_of_pcap}) as in
Eq.~(\ref{Pacc}):
\begin{equation}
P_{\rm cap,high1}
 = \frac{12(12 f_{\rm c}\xi)^{1/\alpha}}
        {\pi \tilde{i} (\tilde{e}^2+\tilde{i}^2)^\gamma}
   R(k,\gamma),
\end{equation}
where $\gamma \equiv (\alpha+2)/2\alpha$, and $R(k,\gamma)$ is an
integral similar to the complete elliptic integral, and is defined by
\begin{equation}
R(k,\gamma)
 \equiv \int_0^1
            \frac{dx}
                 {(1-x^2)^{1/2}(1-k^2x^2)^\gamma}.
\end{equation}
The Rayleigh distribution average of $P_{\rm cap,high1}$ can be
calculated as
\begin{equation}
\ave{P_{\rm cap}}_{\rm high1}
\simeq
   \frac{24(12 f_{\rm c}\xi)^{1/\alpha} R(k,\gamma)}
        {\pi (\tilde{e}^\ast)^{2\gamma-1} (\tilde{i}^\ast)^2}
   \Gamma\left( \frac{3}{2} - \gamma \right)
   {_2F_1} \left( \frac{1}{2},
                  \frac{3}{2} - \gamma,
                  \frac{3}{2},
                  1-\frac{(\tilde{e}^\ast)^2}
                         {(\tilde{i}^\ast)^2}
           \right),
\label{Pcap_high1_ave}
\end{equation}
where $\Gamma$ is the gamma function, $_2F_1$ is the hypergeometric
function, and we assume $\tilde{e}^\ast > \tilde{i}^\ast$ and $\alpha >
1$.
This gives $\ave{P_{\rm cap}}_{\rm high1} \propto
(\tilde{e}^\ast)^{-(2\gamma+1)} =
(\tilde{e}^\ast)^{-2(\alpha+1)/\alpha}$ when $\tilde{e}^\ast =
2\tilde{i}^\ast$, whose dependence on $\tilde{e}^\ast$ (i.e., $\propto
(\tilde{e}^\ast)^{-2(\alpha+1)/\alpha}$) agrees with that of $P_{\rm
cap,high1}$ on $\tilde{e}$ when $\tilde{e}=2\tilde{i}$.  For the typical
case of $\alpha=3$ and $\tilde{e}^\ast=2\tilde{i}^\ast$, we have
$\ave{P_{\rm cap}}_{\rm high1} \simeq 1.52\e{2} (f_{\rm c}\xi)^{1/3}
(\tilde{e}^\ast)^{-8/3}$.
In deriving Eq.~(\ref{Pcap_high1_ave}), we treated $R(k,\gamma)$ as
a constant\footnote{We confirmed that the error arisen from this
simplification is about 3\% when $e^\ast = 2i^\ast$ and $\alpha=3$.},
because $R(k,\gamma)$ is a slowly varying function of $k$ as long as $k$
is not close to unity (note that $0\leq k \leq \sqrt{3/4}$ from its
definition).  For the still higher velocity regime,
\begin{equation}
P_{\rm cap,high2}
 = \frac{2(2 f_{\rm c}\xi)^{2/(\alpha-1)}}{\pi}
   \sqrt{(\tilde{e}/\tilde{i})^2+1}
   E(k),
\end{equation}
thus
\begin{equation}
\ave{P_{\rm cap}}_{\rm high2}
 = \frac{(2 f_{\rm c}\xi)^{2/(\alpha-1)}}{2\pi}
   {\cal F}(I^\ast).
\end{equation}

The curves in Fig.~\ref{fig_Pacc_ave} show $\ave{P_{\rm
acc}}(\tilde{r}_{\rm p}, \xi, \alpha, \tilde{e}^\ast, \tilde{i}^\ast)$
obtained using the above semi-analytic expressions.  We can see that
$\ave{P_{\rm acc}}$ is well approximated by the above semi-analytical
expressions within a factor of two for the entire regions of random
velocity and $\xi$.  When $\tilde{r}_{\rm p}$ is large, $\ave{P_{\rm
acc}}$ tends to approach $\ave{P_{\rm col}}$ because planetesimals tend
to collide with the planet more easily before losing a sufficient amount
of energy by gas drag to get captured.  For example, when
$\tilde{r}_{\rm p} = 0.005$ (left panel) and $\xi=10^{-9}$, $\ave{P_{\rm
acc}} = \ave{P_{\rm col}}$, which can be deduced by Fig.~\ref{fig_R_c}.
On the other hand, $\ave{P_{\rm acc}}$ can be approximated by
$\ave{P_{\rm cap}}$ when $\xi$ is large.
In summary, the effect of atmosphere is statistically negligible in the
entire velocity regime when $\xi$ is so small that $\ave{P_{\rm
col}}_{\rm high} > \ave{P_{\rm cap}}_{\rm high1}$ is satisfied, where
$\ave{P_{\rm col}}_{\rm high}$ and $\ave{P_{\rm cap}}_{\rm high1}$ are
given by Eqs.(\ref{Pcol_ave_high}) and (\ref{Pcap_high1_ave}),
respectively.  This condition can be written in terms of $\xi$ as
\begin{equation}
\xi < 4\e{-9}
      \kakkoi{\tilde{r}_{\rm p}}{0.005}{3}.
\end{equation}
To derive the condition, we set $\tilde{e}^\ast = 2$ as the lowest limit
of the range for the functions.

It should be noted that when the mass of planetesimals are small
($\lesssim 10^{14}$g), the assumption of $\tilde{e}^\ast =
2\tilde{i}^\ast$ breaks down if protoplanetary gas disk still exists and
runaway bodies significantly perturb the orbit of such small bodies
\citep{Ohtsuki2002}.  This is because, once inclination is damped to
$\tilde{i} \leq 1$, the stirring rate of inclination by the large bodies
is much smaller than that of eccentricity in such a velocity regime, and
only eccentricity tends to be enhanced \citep{Ida1990}.

Using analytic calculation of planetesimals' energy dissipation due to
gas drag under the two-body approximation neglecting the solar gravity,
\citet{II03} derived a relation between the size of planetesimals and
the enhanced radius of protoplanets, for a given atmospheric structure
(their Eq.(17)).  Substituting this enhanced radius for the planetary
radius in the expressions for collision rates derived by
\citet{Inaba2001}, semi-analytic expressions of accretion rates for
protoplanets with atmospheres can be obtained \citep[their
Eqs.(20)-(24)]{II03}.
We find that $\ave{P_{\rm acc}}$ with the effect of atmospheric gas drag
obtained from these expressions derived by \citet{II03} ($\ave{P_{\rm
col}}$ in their notation) is smaller by a factor of about two as
compared to ours in the high-velocity regime, even when the same
atmospheric structure is assumed (Fig.\ref{fig_Compare_with_II03}).
This discrepancy is caused by the more simplified procedures in deriving
the expression in \citet{II03} as compared to ours; the cause for the
discrepancy is further discussed in detail in Appendix B.
On the other hand, \citet{II03} obtained the enhanced radius using more
realistic atmospheric structures, while we assumed a power-law density
distribution.  Effects of this assumption will be discussed in Section
\ref{sec_realistic_atmosphere}.

\section{Comparison with the case of more realistic atmospheric
  structures}
\label{sec_realistic_atmosphere}
Here, we examine the validity of our assumption of the simplified
atmospheric structure, i.e., power-law function (Eq.(\ref{rho_g})).  As
shown above, the assumption of power-law approximation for gas density
in atmospheres allowed a systematic study and understanding of the
dependence of the accretion rate on various physical parameters.  In
order to check the validity of this assumption, we compared the results
shown previously in this paper with those obtained using a more
realistic atmospheric structure based on the analytic solution shown by
\citet{II03}.
Although the index $\alpha$ should be properly chosen depending on the
parameter values for a given atmospheric structure, we found that the
use of power law functions is a good approximation in most cases to
describe realistic atmospheres when opacity depletion factor $f_\kappa$
is much smaller than unity ($f_\kappa \lesssim 10^{-2}$), because
optically thin atmospheres tend to be radiative rather than convective.
The assumption of small $f_\kappa$ would be reasonable, because a large
fraction of dust in protoplanetary disks in this stage would be depleted
after the accretion of large bodies such as planetesimals or
protoplanets.
In addition, small $f_\kappa$ also corresponds to the situation when
dust size is much larger than sub-micron size, and this would be the
case for large protoplanets ($M\gtrsim M_\oplus$), because dust in the
atmospheres grows quickly to reduce opacity by a factor of 100 from that
of sub-micron size dust \citep{Movshovitz2008}.
%

However, the power-law distribution would not be a good approximation
at outer region of the atmospheres in some cases.
The density profile fitted by a power-law function at the bottom of an
atmosphere tends to underestimate the gas density at the outer region
when planet mass is small ($M \sim 0.1M_\oplus$) and overestimate it
when planet mass is large ($M \sim 10M_\oplus$) for typical disk
parameters.
In the case of small planets, however, the atmosphere is not very
effective in capturing planetesimals because it is not massive enough.
On the other hand, the effect of atmospheric gas drag on accretion rates
is most important for such large bodies as $M \sim 10M_\oplus$, because
they have thick and massive atmospheres \citep{II03}.  Also, the
formation timescale of protoplanets is determined by the later stage of
accretion (i.e., when $M \sim$ 1--10$M_\oplus$ for $\sim$ 5AU) where
their growth timescale is long.  Thus here we concentrate on the cases
of large protoplanets, and compare accretion rates for atmospheres with
power-law density profiles and those for more realistic atmospheric
structures.
As examples, we consider cases of $1M_\oplus$ and $10M_\oplus$ planets
at 5AU when disk gas temperature is 125K and density is $1.5\e{-8}{\rm
kg\,m^{-3}}$.  We set the density of solid part of the planets to be
$3.4\e{3}{\rm kg\, m^{-3}}$ ($\tilde{r}_{\rm p} = 0.001$), which
represents the density of $10M_\oplus$ planets that consist of ice and
rock for roughly even mass \citep[e.g.,][]{Fortney2007}.

Figure \ref{fig_Ikoma-atmosphere} shows the density profiles derived by
the method given by \citet{II03} together with fitted power-law
profiles.  As we can see, the fitted power-law functions overestimate
the gas density in the outer region $(\tilde{r}\gtrsim 0.01)$, by a
factor of up to $\sim 10^2$ for the $10M_\oplus$ case.  As for the
$1M_\oplus$ case, the line for the power-law distribution crosses the
realistic one and underestimates the gas density at the outermost region
$(\tilde{r} > 0.2)$ because the region is already outside of the outer
boundary of the analytic solution given by \citet{II03} and the
temperature and density are constant.  The density profile at the outer
region also depends on the outer boundary condition, which will be
briefly discussed later in this section.

Figure \ref{fig_DeltaE-rmin_for_ikoma-atm} shows $\Delta \tilde{E}$
obtained by orbital integration as a function of $\tilde{r}_{\rm min}$
for the two density profiles for the $10M_\oplus$ planet for various
planetesimal sizes ($10$ -- $10^7$m) with $\tilde{e}= 0.1$ and 10 in
order to examine the dependence of $\tilde{R}_{\rm c}$ on $r_{\rm s}$
(or $\xi$) and $\tilde{e}$ when $\tilde{e}=2\tilde{i}$ (the values of
$\tilde{r}_{\rm min}$ for the upper end of the horizontal band for each
curve in this plot corresponds to $\tilde{R}_{\rm c}$, as we discussed
in \S \ref{sec_gas_drag}).
In the case of $\tilde{e}=0.1$ (left panel), we can see from the shapes
of the curves that the difference of $\Delta \tilde{E}$ directly
reflects the assumed density profiles (Fig.~\ref{fig_Ikoma-atmosphere}).
When planetesimals are large ($r_{\rm s}=10^4$km), $\tilde{R}_{\rm c}$
of the two atmospheres are almost the same because the two curves
eventually agree where the curves reach $\Delta \tilde{E} \sim 2-3$,
which is necessary to be captured within the Hill sphere when
$\tilde{e}\sim 0.1$.  But as planetesimals become smaller, the curves of
$\Delta \tilde{E}$, which is dissipation energy per unit mass, shift
upward, which yields two different $\tilde{R}_{\rm c}$, the values for
the realistic atmospheric model being smaller.  When $r_{\rm s} =$1km,
the difference in the value of $\tilde{R}_{\rm c}$ for the two
atmospheric profiles is about a factor of three.
In the case of $\tilde{e}=10$ (right panel), energy dissipation required
for capture is larger.  Thus, in order to be captured, planetesimals
have to pass through the inner dense region of the atmosphere, where the
difference of the two density profiles are smaller, yielding better
agreement in $\tilde{R}_{\rm c}$ for $\tilde{r}_{\rm s} \geq 1$km.

Figure \ref{fig_pacc_for_ikoma-atm} shows the plots of $P_{\rm acc}$ in
the case of $\tilde{e} = 2\tilde{i}$ for the two atmospheres for various
planetesimal sizes.
In the case of $M=10M_\oplus$ (left panel), $P_{\rm acc}$ for the fitted
power-law density profile overestimates the accretion rate for small
planetesimal sizes, as easily deduced from the results of
Figs.~\ref{fig_Ikoma-atmosphere} or \ref{fig_DeltaE-rmin_for_ikoma-atm},
whereas the two values of $P_{\rm acc}$ for the two atmospheric models
are almost the same when planetesimals are large.  Smaller planetesimal
cases tend to give a larger difference in $P_{\rm acc}$ because
planetesimals are captured at outer atmospheres, where the difference of
the two density profiles is larger.  The values of $P_{\rm acc}$ differ
by a factor of 2--3 for 1km planetesimals, and the difference is about a
factor of 15 for 10m planetesimals.
As for the dependence on eccentricity, there are two effects that shift
$P_{\rm acc}$ in opposite directions.
One is the $\tilde{e}$-dependence of the amount of energy dissipation
required for capture, which we already discussed above and the
difference in $P_{\rm acc}$ for the two atmospheric profiles becomes
smaller for planetesimals with higher eccentricities
(Fig.~\ref{fig_DeltaE-rmin_for_ikoma-atm}).  This tendency can be seen
in the case of $r_{\rm s}=10^2$km, where the difference between the
results for the two atmospheric profiles becomes smaller with increasing
$\tilde{e}$.
The other is the effect of gravitational focusing by the planet.  When
the initial random velocity of incoming planetesimals is small
($\tilde{e} \lesssim 1$), velocities of planetesimals at $\tilde{R}_{\rm
c}$ ($\ll 1$) are determined by the planet's gravity, and thus the
effective collision cross section is proportional to $\tilde{R}_{\rm c}$
({\it cf.} Eq.~(\ref{integrand_of_pcap})).  Therefore\footnote{Note that
$P_{\rm acc}$ is a quantity integrated with respect to $\tilde{b}$,
which means that $P_{\rm acc}$ is not directly proportional to
$\tilde{R}_{\rm c}$.  But the dependence of $p_{\rm col}$ or $p_{\rm
cap}$ on $\tilde{b}$ is not large, as shown in Fig.~\ref{fig_dp-cap}.},
the difference in $P_{\rm acc}$ is roughly proportional to the
difference of $\tilde{R}_{\rm c}$.  However, when the initial random
velocity is large, planetesimals are not significantly accelerated by
the planet's gravity at locations with $\tilde{r} \gtrsim
6/\tilde{e}^2$.  Thus, when $\tilde{R}_{\rm c} \gtrsim 6/\tilde{e}^2$,
$P_{\rm acc}$ tends to be proportional to the geometrical cross section,
i.e., $\propto \tilde{R}_{\rm c}^2$, which amplifies the difference in
$P_{\rm acc}$.
\footnote{Note that such a high velocity, which results in the large
difference in $P_{\rm acc}$ for planetesimals with $r_{\rm s}=10$m size
(by a factor of 15), is unlikely to be realized.  Because, to enhance
the relative velocity such that $P_{\rm acc}$ is roughly $\propto$
$\tilde{R}_{\rm c}^2$, planetesimals need a close encounter that minimum
distance to the protoplanet have to be closer to the enhanced radius,
which means that planetesimals would be captured by the encounter.}
This tendency can be seen in the case of $r_{\rm s}=10$m, where the
difference between the results for the two atmospheric profiles becomes
larger with increasing $\tilde{e}$.
In the case of $M=1M_\oplus$, the difference between the values of
$P_{\rm acc}$ for the two atmospheric profiles can be explained
basically in a similar way.  In both $1M_\oplus$ and $10M_\oplus$ cases,
the difference between the results for the atmospheric profiles is
rather small in the low-velocity regime with $\tilde{e} \leq 1$.  One
difference from the $M=10M_\oplus$ case is that $P_{\rm acc}$ with the
power-law model for the smallest planetesimal case ($r_{\rm s}=0.1$m)
underestimates the accretion rate when $\tilde{e}\lesssim 10$, which
simply reflects the underestimation of the density at the outer region
(see Fig.~\ref{fig_Ikoma-atmosphere}).
However, it should be noted that the assumption of our numerical method
(i.e., planetesimals move on a Keplerian orbit while weakly perturbed by
gas drag) breaks down for such small planetesimals.  Their motion is
expected to be strongly coupled with the gas flow, and the accretion
rate of such small planetesimals is determined by the flow pattern of
the nebular gas in the vicinity of a planet, rather than the structure
of the atmosphere near the planet's surface.  Therefore, the difference
in the accretion rates for the two atmospheric profiles is practically
negligible in the $M=1M_\oplus$ case, if we take into account the fact
that sufficiently small planetesimals are coupled with the gas drag flow.

In summary, the power-law atmospheric density profile is a reasonable
approximation for accretion rates of large ($\sim 10M_\oplus$)
protoplanets when planetesimals' size is not too small.
In the case of $1M_\oplus$, the difference of the two atmospheric
profiles practically does not cause a significant difference in $P_{\rm
acc}$,
if we take account of the fact that the motion of planetesimals with
very small sizes ($\lesssim$ 0.1m) are strongly coupled with the nebular
gas.

We further note that the outer boundary condition of atmospheres is
important for the structure of the outer region of atmospheres, where
the difference between realistic density profiles and fitted power-law
profiles can be significant, although the inner dense region, which
account for most of the atmospheric mass, is insensitive to the
boundary condition.
For example, when the temperature of the outer boundary is decreased to
50K, which is a typical disk temperature at 5AU for a disk that is
optically thin in the vertical direction and optically thick in the
horizontal direction, the agreement between the realistic density
profile and the fitted power-law function is significantly improved (see
Fig.~\ref{fig_Ikoma-atmosphere}); in this case, the difference in
accretion rates for the two atmospheric profiles is expected to be
reduced greatly.  Conversely, when the density at the outer boundary
decreases due to the dispersal of the nebular gas, for example, the
difference between the two profile increases
(Fig.~\ref{fig_Ikoma-atmosphere}).

\section{Growth rates of protoplanets}\label{sec_growth}
In order to examine the consequence of the obtained formula of accretion
rates, we calculate the time evolution of planet mass using a simple
semi-analytic model with the accretion rate obtained above.  We
basically follow the calculation model of \citet{Chambers2006a}, and
also compare our results with his results.

We consider a protoplanet embedded in a disk of single-size
planetesimals, whose random velocity is determined by the equilibrium
between damping by gas drag and excitation by the protoplanet.  The
growth rate of a protoplanet is given by
\begin{equation}
\frac{dM}{dt}
 = \left( \frac{2\pi \Sigma_{\rm p}r_{\rm H}^2}{T}
   \right)
   \ave{P_{\rm acc}},
\label{dMdt}
\end{equation}
where $\Sigma_{\rm p}$ is the surface density of planetesimals and $T$
is the orbital period of the protoplanet.  To integrate this equation,
we need formulas for the surface density and $\ave{P_{\rm acc}}$.  The
surface density is given by
\begin{equation}
\Sigma_{\rm p}
 = \Sigma_{\rm tot} - \frac{M}{2\pi a \tilde{b}r_{\rm H}},
\end{equation}
where $\Sigma_{\rm tot}$ is the initial surface density of
planetesimals.  This expression was derived under the assumption that
the protoplanet does not migrate radially.  In this case, the final mass
of the planet becomes the isolation mass, $(2\pi a^2 \tilde{b}_{\rm f}
\Sigma_{\rm tot})^{3/2}/(3M_\ast)^{1/2}$, where $\tilde{b}_{\rm f}$ is
the width of the feeding zone of the planet scaled by the Hill radius.
As for the normalized accretion rate, we use our semi-analytic
expression for the high-velocity regime (Eq.(\ref{Pcap_high1_ave})) in
accordance with \citet{Chambers2006a}, who used the expression based on
\citet{II03} for the same velocity regime.  Equation
(\ref{Pcap_high1_ave}) includes the gas drag parameter $\xi$, which is a
function of the accretion rate (see Eq.(\ref{xi_typical})).  Thus we
solve simultaneous equations with respect to $dM/dt$ (Eq.(\ref{dMdt})
with Eq.(\ref{Pcap_high1_ave})) and $\xi$ self-consistently.
We set $\Sigma_{\rm tot} = 100\, {\rm kg\,m^{-2}}$ and $a=5$AU, and also
assume that mean molecular weight is 2.8, material density is $2.0\e{3}\,
{\rm kg\,m^{-3}}$, gas density of the disk is $3.9\e{-8} {\rm
kg\,m^{-3}}$, opacity is $1.0\e{-3} {\rm m^2\,kg^{-1}}$, and $\alpha=3$
for the atmospheric gas density distribution.

Figure \ref{fig_Chambers2006_Fig8_compare}(a) shows the growth of the
protoplanet.  We first calculated the growth with the formula used in
\citet{Chambers2006a} to check our calculation, and confirmed that their
results in the case of planetesimals with 10km diameter can be
reproduced (thin dashed line in
Fig.~\ref{fig_Chambers2006_Fig8_compare}(a); see also Fig.8 of
\citet{Chambers2006a}).
As demonstrated by the semi-analytic calculation of
\citet{Chambers2006a}, acceleration of the growth of protoplanets by the
effect of atmospheric gas drag becomes notable when $M>M_\oplus$,
because at this stage the atmospheres of the protoplanets become dense
enough to capture 10-km-diameter planetesimals by gas drag.

On the other hand, the three thick lines in
Fig.~\ref{fig_Chambers2006_Fig8_compare}(a) show results obtained using
our new formula for accretion rates with the effect of atmospheric gas
drag.  We also plot the change of values of $\xi$ in the course of
planetary growth in these three cases in
Fig.~\ref{fig_Chambers2006_Fig8_compare}(b).
We can see that the growth based on our formula of the accretion rate
for planetesimals with 10km is faster than that obtained by
\citet{Chambers2006a} by a factor of about two, which reflects the
difference between our formula and that of \citet{II03}, as discussed in
\S \ref{sec_empirical_formula} and Appendix B.
In addition to the case of 10-km-diameter planetesimals, we also
performed calculations with smaller planetesimal sizes (1km and 100m).
The acceleration of the protoplanet's growth is much more significant in
the case of small planetesimals for two reasons.
First, in the early stage where $M \lesssim 10^{-2}M_\oplus$, the growth
rate is enhanced for the small planetesimal cases because their
equilibrium random velocity is smaller due to strong drag by the
protoplanetary disk gas than that in the case of the larger planetesimal
case.
Second, in the late stage where $M \gtrsim 0.1M_\oplus$, the enhancement
of accretion rates by atmospheric gas drag is more significant in the
small planetesimal case, because corresponding values of $\xi$ are
larger for small planetesimals as shown in
Fig.~\ref{fig_Chambers2006_Fig8_compare}(b).  As a result, the growth is
accelerated by more than a factor of 10 in the case of 100-m-diameter
planetesimals as compared to the case of 10-km-diameter.
We do not intend to draw many conclusions from our growth calculation
presented in Fig.~\ref{fig_Chambers2006_Fig8_compare}, because our model
is rather simple and does not include other important effects such as
fragmentation or migration.  However, our calculation shows the
importance of atmospheric gas drag in planetary accretion, as first
demonstrated and emphasized by \citet{II03} and \citet{Inaba_etal_2003},
and it also demonstrates the usefulness of the new formula for accretion
rates derived in the present work.
A more detailed growth model, which includes fragmentation of
planetesimals, non-equilibrium eccentricity, and radial migration as
well as planetesimal capture by atmospheric gas drag can be found in
\citet{Chambers2006a,Chambers2008}.

\section{Conclusions and Discussion}\label{sec_summary}
In the present work, we examined accretion rates of planetesimals onto
protoplanets that have atmospheres by analytic calculation and numerical
orbital integration with gas drag.
Assuming that the radial distribution of the atmospheric gas density can
be approximated by a power-law and that most of energy dissipation of a
planetesimal passing through a protoplanet's atmosphere occurs near the
point of closest approach to the protoplanet, we analytically estimated
dissipation of kinetic energy of the planetesimal, and confirmed
agreement with orbital integration.  We performed three-body calculation
for a large number of orbits and obtained accretion rates for a wide
range of parameters.  We also examined the case when planetesimals have
the Rayleigh distribution in orbital eccentricities and inclinations,
and derived a semi-analytic expression of the accretion rate with the
effect of atmospheric gas drag.

We found that our semi-analytic formula given by Eq.~(\ref{Pacc_ave})
can reproduce our numerical results for the accretion rate of
planetesimals with the Rayleigh distribution in eccentricities and
inclinations by planets with atmospheres very well.  The formula is
basically described as a function of five non-dimensional parameters:
the normalized radius of solid surface of planets $\tilde{r}_{\rm p}$,
the non-dimensional gas-drag coefficient $\xi$, the exponent of
power-law function for the gas density profile $\alpha$, and
r.m.s.~eccentricity $\tilde{e}^\ast$ and inclination $\tilde{i}^\ast$ of
planetesimals.  Many other physical parameters with real dimension such
as semi-major axis, planet mass, radius of incoming particles, etc. are
all reduced to $\xi$ as in Eq.(\ref{xi_typical}), thus our results can
be applied to a wide range of situations through the above
non-dimensional parameters.

We also performed orbital calculation with gas drag from an atmosphere
with a realistic density distribution, and compared with the above
results for an atmosphere with a power-law density distribution.  We
found that the results using these two different atmospheric profiles
agree well with each other, for example, when the protoplanet's mass is
$10M_\oplus$ and planetesimals are not too small ($r_{\rm s}\gtrsim
1$km), while the results using the power-law density distribution tends
to overestimate the accretion rate when planetesimals' size is smaller
and their random velocity is large ($\tilde{e} \gg 1$). We also found
that the degree of deviation depends on the outer boundary condition of
the atmosphere because the structure of the outer atmosphere is
sensitive to the boundary.

Using a simple semi-analytic model and the above results of accretion
rate with atmospheric gas drag, we performed calculation for the growth
of protoplanets.  We confirmed the results of previous studies that
atmospheric gas drag can significantly enhance the growth rate of
protoplanets \citep{II03,Inaba_etal_2003,Chambers2006a}.  We also found
that the acceleration of the growth is significantly enhanced for small
planetesimals.  Since our semi-analytic formula is expressed in a
general form in terms of non-dimensional parameters, it can be used in
the study of planetary accretion in various situations.



\begin{center}
 {\bf Acknowledgments}
\end{center}

We thank H. Tanaka, H. Kobayashi, and S. Inaba for helpful comments.
T.T. is also grateful to S. Ida for fruitful discussion and continuous
encouragement.  We also thank J. Chambers for useful comments that led
to an improvement of the paper.
This work was supported by NASA's Origins of Solar Systems Program
(NNG05GH87G, NNX08AI37G),
the Ministry of Education, Culture, Sports, Science and Technology of
Japan (MEXT) Grant-in-Aid for Scientific Research on Priority Areas,
``Development of Extrasolar Planetary Science'' (MEXT-16077202),
and CPS running under the auspices of the MEXT Global COE Program
entitled ``Foundation of International Center for Planetary Science''.
Numerical simulation was performed at the Global Scientific Information
and Computing Center of Tokyo Institute of Technology and at the Center
for Planning and Information Systems of Japan Aerospace Exploration
Agency.






\vspace{5mm}
\begin{center}
 {\bf Appendix A: Gas density distribution of a wholly radiative
 atmosphere near the planetary surface}
\end{center}
Equations of hydrostatic equilibrium and diffusion approximation of
radiative transfer are given by
\begin{equation}
\frac{dP}{dr}
 = -\frac{GM}{r^2} \rho,
\label{hydrostatic_eq}
\end{equation}
\begin{equation}
\frac{16\sigma T^3}{3\kappa\rho}
\frac{dT}{dr}
 = -\frac{L}{4\pi r^2},
\label{radiative_dif_approx}
\end{equation}
where $P$ is pressure, $G$ is the gravitational constant, $M$ is
planet's mass, $r$ is distance from the planet center, $\rho$ is gas
density, $\sigma$ is the Stefan-Boltzmann constant, $T$ is gas
temperature, $L$ is luminosity, and $\kappa$ is opacity.  Eliminating
$r$ from equations (\ref{hydrostatic_eq}) and
(\ref{radiative_dif_approx}), we have
\begin{equation}
\frac{dP}{dT}
 = \frac{64\pi \sigma GM}{3\kappa L} T^3.
\end{equation}
Integrating this equation, we have
\begin{equation}
P
 = \frac{16\pi \sigma GM}{3\kappa L} T^4,
\end{equation}
where temperature and pressure at the outer boundary (photosphere) is
assumed to be much smaller than those of atmospheres near the planetary
surface.  Using the equation of state $P = \rho R' T$ ($R'$ is the gas
constant),
\begin{equation}
\rho
 = \frac{16\pi \sigma GM}{3R'\kappa L} T^3.
\label{rho_radiative}
\end{equation}
Substitution of this into Eq.~(\ref{radiative_dif_approx}) yields
\begin{equation}
\frac{dT}{dr}
 = -\frac{GM}{4R'r^2}.
\end{equation}
Integrating this, we have
\begin{equation}
T = \frac{GM}{4R'r}.
\end{equation}
Substituting this into Eq.~(\ref{rho_radiative}), we have
\begin{equation}
\rho
 = \frac{\pi \sigma G^4 M^4}
        {12 R'^4 \kappa L r^3}.
\end{equation}
Since $M=(4/3)\pi r_{\rm p}^3 \rho_{\rm core}$, where $r_{\rm p}$ and
$\rho_{\rm core}$ are radius and mean density of the planet,
respectively, the density at the surface of the planet can be written in
the form:
\begin{equation}
\rho(r=r_{\rm p})
 = \frac{\pi^2 \sigma \rho_{\rm core} G^4 M^3}
        {9 R'^4 \kappa L}
 \propto M^{3} L^{-1} \kappa^{-1}.
\end{equation}

\begin{figure}
\epsscale{0.5}
\plotone{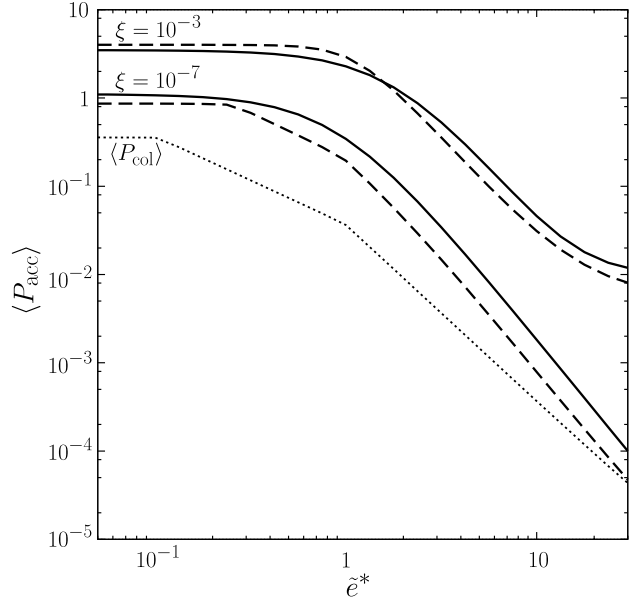}
\caption{ Normalized accretion rate averaged over the Rayleigh velocity
distribution as a function of $\tilde{e}^\ast (=2\tilde{i}^\ast)$ for
$\xi = 10^{-7}$ and $\tilde{r}_{\rm p}=0.001$.  Solid line shows the
accretion rate obtained by our semi-analytic expression
(Eq.(\ref{Pacc_ave})), and dashed line is obtained by using the formulas
derived by \citet{II03}.  In both cases, a power-law density
distribution is assumed for the atmospheric structure ($\alpha=3$).
Collision rate for the case without an atmosphere is also shown by the
dotted line for comparison.
\label{fig_Compare_with_II03}}
\end{figure}

\begin{figure}
\epsscale{0.5}
\plotone{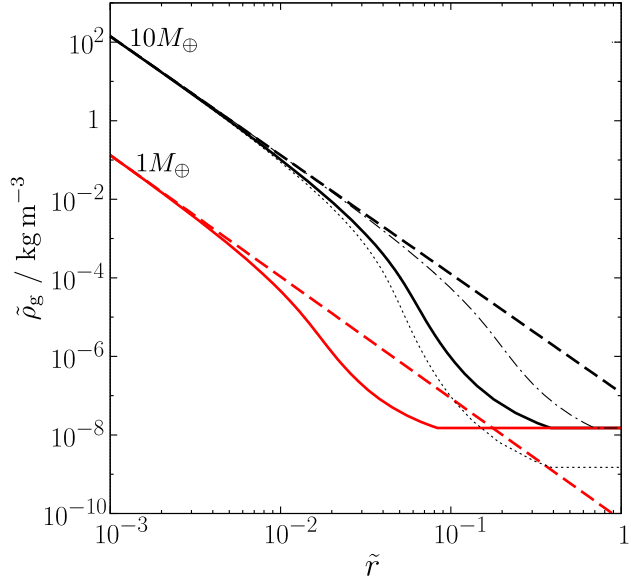}
\caption{Density profiles of a realistic atmosphere derived by the
analytic method given in \citet{II03} for $1M_\oplus$ and $10M_\oplus$
planets at 5AU (thick-solid lines) in the case when $\rho_{\rm
core}=3.4\e{3}{\rm kg\,m^{-3}}$, $L=10^{-7}L_\odot$, and
$f_\kappa=10^{-2}$, and power-law functions fitted to the realistic one
at the bottom (thick-dashed lines).  The exponents for the fit lines are
$\alpha=3.094$ for $1M_\oplus$ and $\alpha=3.027$ for $10M_\oplus$.
Thick lines show the case when the temperature and gas density at the
outer boundary are 125K and $1.5\e{-8}{\rm kg\,m}^{-3}$, respectively.
In order to see the dependence on the boundary condition, we also show
density profiles when the temperature at the boundary is lower (50K,
thin dot-dashed line) and the density is lower ($1.5\e{-9}{\rm
kg\,m^{-3}}$, thin dotted line) in the $10M_\oplus$ case.
\label{fig_Ikoma-atmosphere}}
\end{figure}

\begin{figure}
\epsscale{1.0}
\plottwo{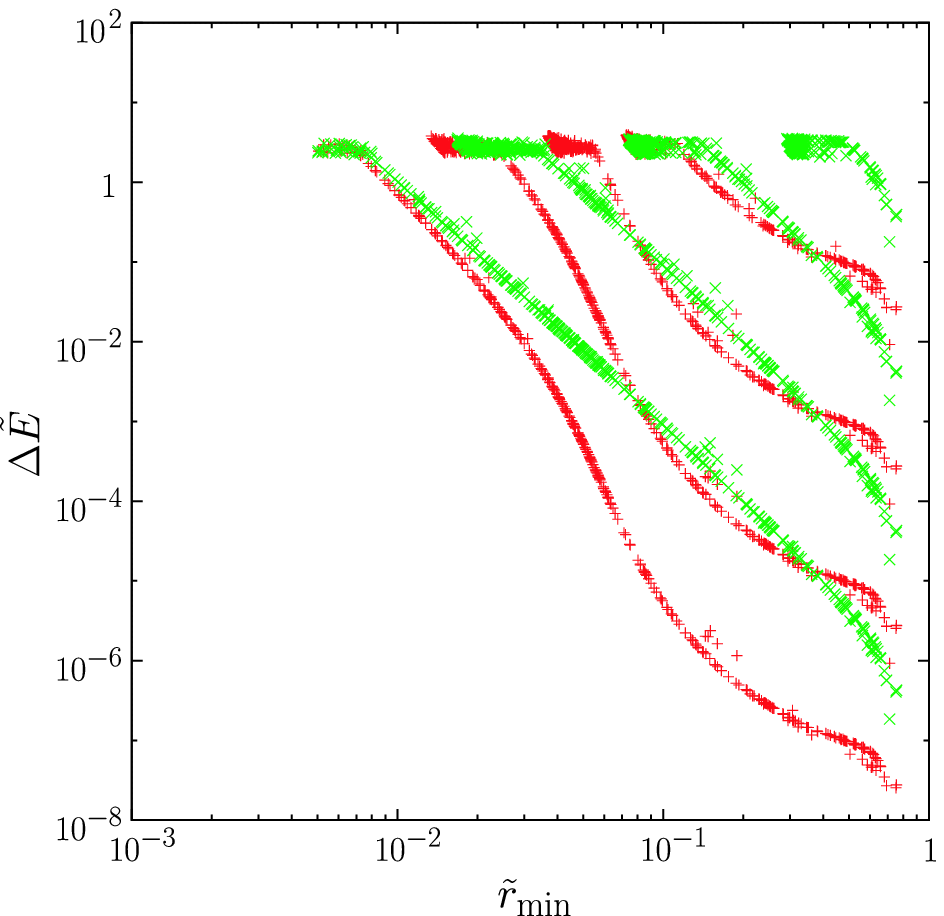}{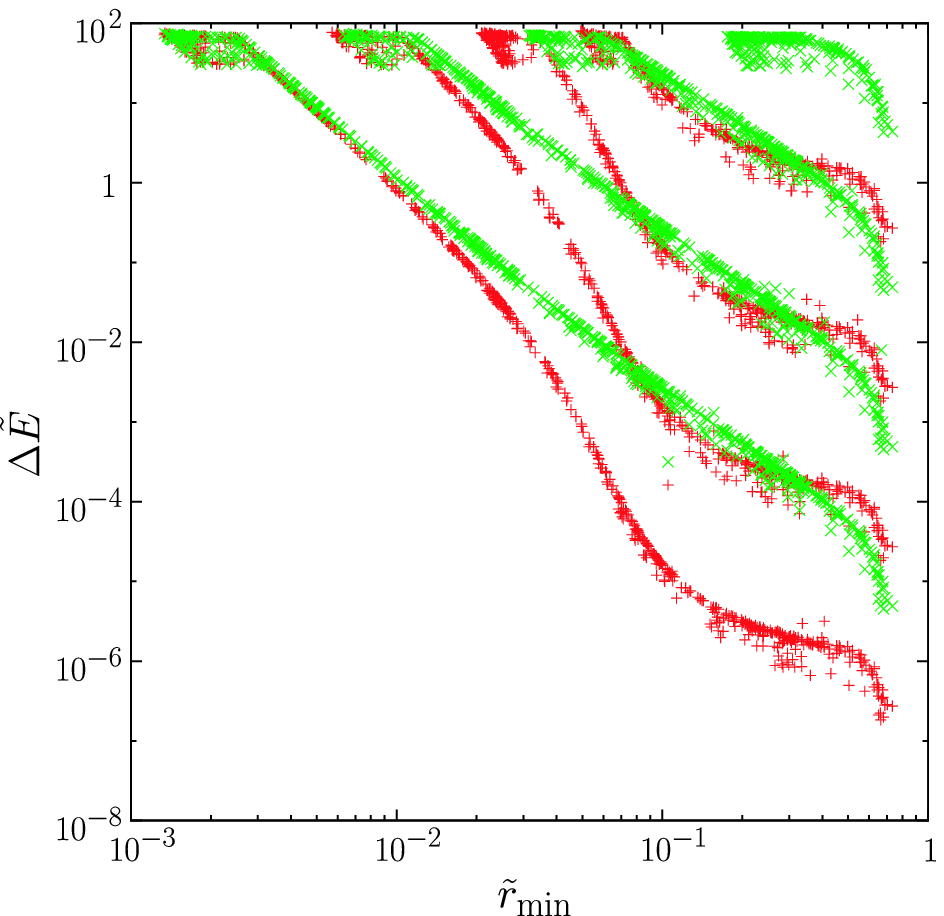}
\caption{Change of the energy $\tilde{E}$ due to gas drag as a function
of minimum approach distance (the same as
Fig.~\ref{fig_Delta-J_xi-e-depend}) for a realistic atmosphere model
(red pluses) and a power-law model (green crosses) when $M=10M_\oplus$
at 5AU.  Left and right panels show the case when $(\tilde{e},\tilde{i})
= (0.1,0.05)$ and $(10,5)$, respectively.  Four different planetesimal
sizes are assumed for each of the two atmosphere models; $r_{\rm
s}=10^4$km, $10^2$km, 1km, and 10m, from lower-left to upper-right.
\label{fig_DeltaE-rmin_for_ikoma-atm}}
\end{figure}

\begin{figure}
\epsscale{1.0}
\plottwo{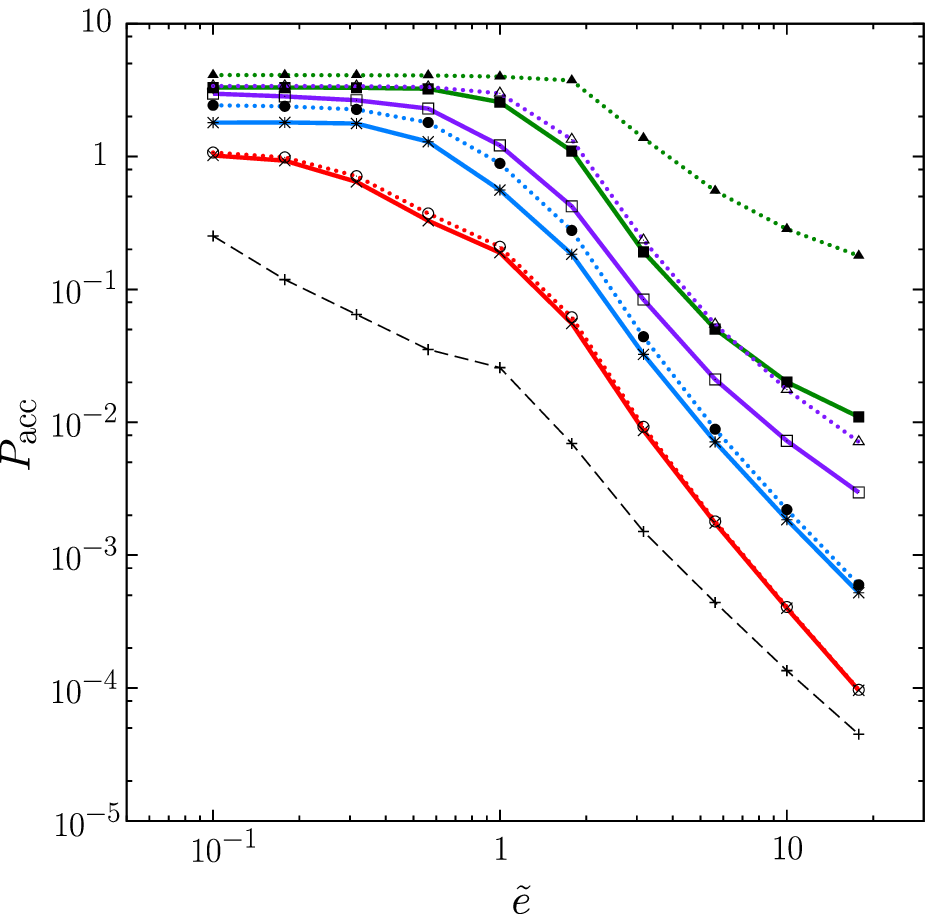}{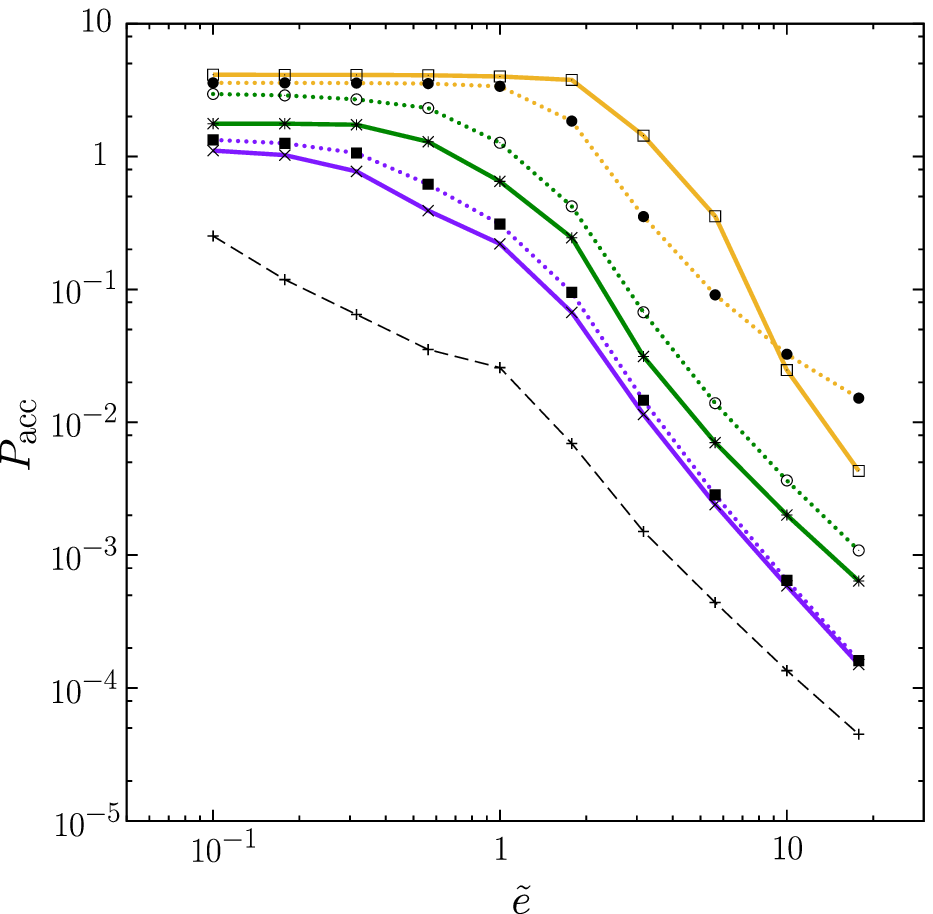}
\caption{$P_{\rm acc}$ for the realistic atmosphere model (thick-solid
lines) and for the power-law atmosphere model (thick-dotted lines), and
$P_{\rm col}$ for $\tilde{r}_{\rm p}=0.001$ (thin dashed line)
($\tilde{e}=2\tilde{i}$).  Left panel shows the cases when
$M=10M_\oplus$ with $r_{\rm s}=10^4$km (red lines), $10^2$km (blue
lines), 1km (purple lines), and 10m (green lines).  Right panel shows
the cases when $M=1M_\oplus$ with $r_{\rm s}=$1km (purple lines), 10m
(green lines), and 0.1m (yellow lines).
\label{fig_pacc_for_ikoma-atm}}
\end{figure}

\begin{figure}
\epsscale{0.9}
\plottwo{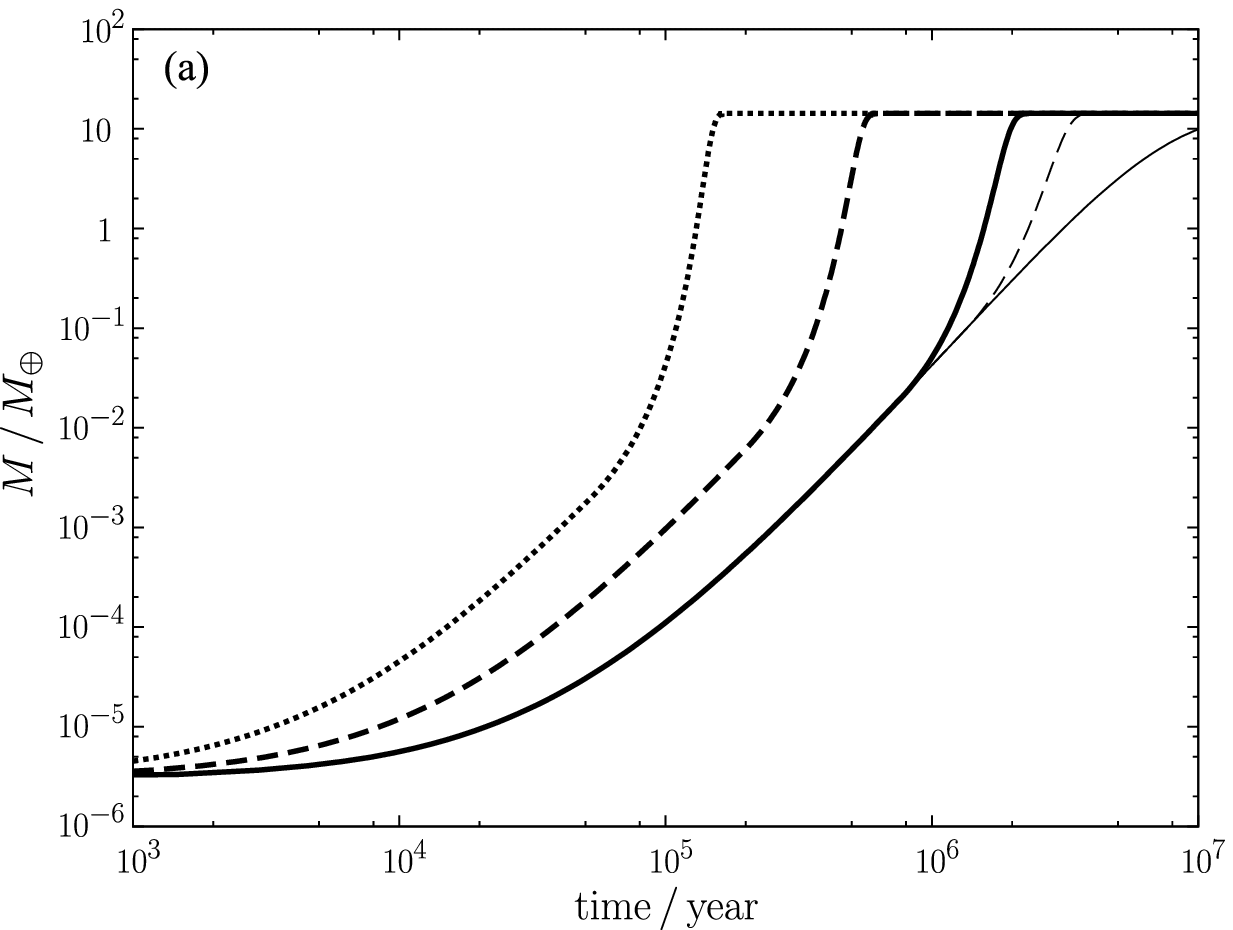}{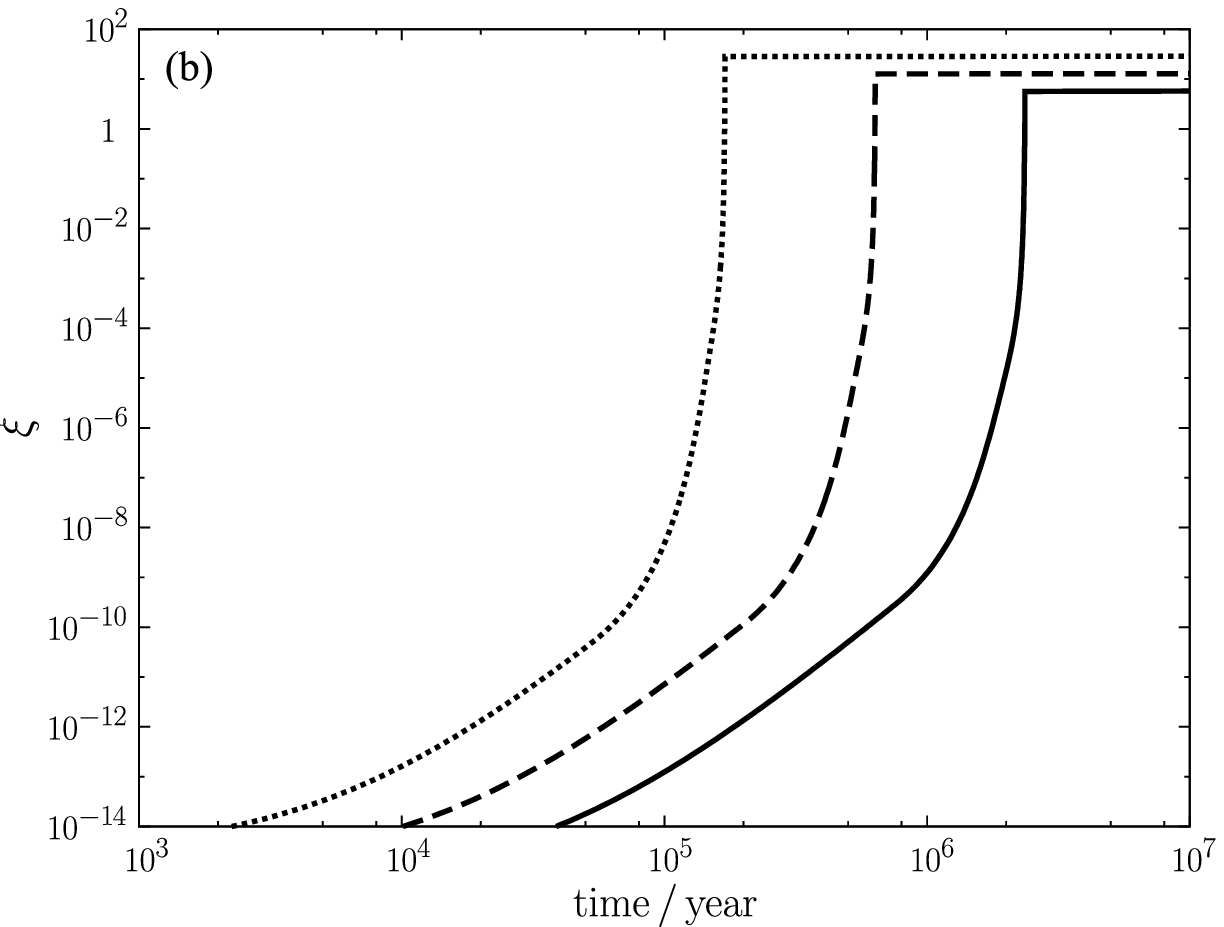}
\caption{(a) Growth of a protoplanet at 5AU from the Sun in a disk with
initial solid surface density of $100 {\rm kg\,m^{-2}}$ and with gas
density of $3.9\e{-8} {\rm kg\,m^{-3}}$.  Planetesimals' material
density is $2\e{3}{\rm kg\,m^{-3}}$.  Thin solid line shows the case of
10-km-diameter planetesimals and a protoplanet without an atmosphere,
and the thin-dashed line shows the case with the effect of atmosphere
based on the formula of \citet{Chambers2006a} for planetesimals of 10km
diameter.  Thick solid, dashed, and dotted lines show the results based
on our formula of the accretion rate with the effect of atmosphere, in
the case with planetesimal of 10km, 1km, and 100m, respectively.  (b)
Time evolution of $\xi$ in the three cases with our formula of accretion
rates for planetesimals with 10km (solid line), 1km (dashed line), and
100m (dotted line), respectively.
\label{fig_Chambers2006_Fig8_compare}}
\end{figure}

\vspace{5mm}
\begin{center}
{\bf Appendix B: Comparison with \citet{II03}}
\end{center}
As shown in Fig.~\ref{fig_Compare_with_II03}, accretion rates derived in
the present work are systematically larger than those obtained by
using the formulas derived by \citet{II03}.
Here, we compare our results given by Eq.~(\ref{Pacc_ave}) with their
results, which are obtained mainly by a simpler way (i.e., two-body
orbital integration and rough estimation of dissipation energy of gas
drag), and we examine the cause of the difference.

Calculating the amount of planetesimals' kinetic energy dissipated by
gas drag under the two-body approximation neglecting the solar gravity,
\citet{II03} derived a relation between the enhanced radius of a
protoplanet and the size of planetesimals, which also depends on the
atmospheric density distribution (their Eq.(17)).  Substituting the
obtained enhanced radius into the semi-analytic expressions of collision
rates derived by \citet{Inaba2001}, accretion rates for protoplanets
with atmospheres are obtained \citep[their Eqs.(20)-(24)]{II03}.

Since $\ave{P_{\rm acc}}$ is a quantity that is integrated with respect
to $\tilde{e}$, $\tilde{i}$, and $\tilde{b}$, $\ave{P_{\rm acc}}$ is not
an explicit function of $\tilde{e}$, $\tilde{i}$, and $\tilde{b}$.
However, $\ave{P_{\rm acc}}$ of \citet{II03} ($\ave{P_{\rm col}}$ in
their notation) is a function of $\tilde{R}_{\rm c}$, which explicitly
depends on $\tilde{e}$, $\tilde{i}$, and $\tilde{b}$.
Here we assume that $\tilde{v}_\infty = \sqrt{(\tilde{e}^\ast)^2 +
(\tilde{i}^\ast)^2}$ for $\tilde{R}_{\rm c}$ in calculating their
$\ave{P_{\rm acc}}$.  In the following comparison, we assume an
atmospheric structure described by a power-law function (i.e.,
Eq.(\ref{rho_g})).

Figure \ref{fig_Compare_with_II03} shows normalized accretion rates
obtained by our semi-analytic expression and that of \citet{II03} when
$\xi=10^{-7}$ and $\tilde{r}_{\rm p}=10^{-3}$.  In this case, all
accreting objects are practically captured by the gas drag of the
atmosphere, rather than direct collision.  We can see that the overall
features of the two results are similar to each other; the accretion
rates in the low velocity regime roughly agree with each other and those
in the high velocity regime have the same slope.  However $\ave{P_{\rm
acc}}$ for $\xi=10^{-7}$ of our result is about a factor of 2.3 higher
than theirs.
%
We find that the difference of the factor of 2.3 between the two
formulas comes from the following three effects:
(i) The amount of energy dissipation for incoming particles in an
atmosphere is estimated by their Eq.(16) in \citet{II03}.  This estimate
is smaller by a factor of about two compared to the more accurate
estimate in the present work.  The correction factor is denoted by
$f_{\rm c}$ in the present work and was estimated empirically to be
1.83.  This factor appears in the expression of the accretion rate as
$\ave{P_{\rm acc}} \propto f_{\rm c}^{1/\alpha}$
(Eq.(\ref{Pcap_high1_ave})), thus it enhances $\ave{P_{\rm acc}}$ by
$\sim 1.83^{1/3} = 1.22$.
(ii) \citet{II03} does not seem to fully take account of the
$\tilde{b}$-dependence of $\tilde{R}_{\rm c}$ in obtaining the accretion
rate averaged over the Rayleigh distribution.  If it is correctly taken
into account, this enhances the accretion rate by 1.17.
(iii) Instead of fully taking account of the $\tilde{e}$- and
$\tilde{i}$-dependence of $\tilde{R}_{\rm c}$ and the averaging over the
Rayleigh distribution, $\tilde{v}_\infty = \sqrt{(\tilde{e}^\ast)^2 +
(\tilde{i}^\ast)^2}$ seems to have been used in their calculation of
$\ave{P_{\rm col}}$.  If we fully take account of this, the accretion
rate is enhanced by about 1.6.
Although all of their calculation procedures are not explicitly written
in \citet{II03}, the product of the above three factors is about 2.3,
which is consistent with the difference observed in
Fig.~\ref{fig_Compare_with_II03}, thus we believe that the difference
comes from the three factors described above.  Based on the
consideration, we can deduce the reason why the two accretion rates in
the low velocity regime roughly agree with each other.  As described
above, two causes out of the three are related to the dependence of
$\tilde{R}_{\rm c}$ on $\tilde{e}, \tilde{i}$ and $\tilde{b}$, while
$\tilde{R}_{\rm c}$ does not depend on them in the low velocity regime.

\clearpage

\clearpage

\begin{thebibliography}{}
\bibitem[Benvenuto and Brunini(2008)]{Benvenuto2008} Benvenuto, O. G.,
 Brunini, A., 2008.  The effects of ablation on the cross section of
 planetary envelopes at capturing planetesimals. \icarus 195, 882-894.
\bibitem[Bodenheimer and Pollack(1986)]{Bodenheimer1986} Bodenheimer, P,
 Pollack, J. B., 1986.  Calculations of the accretion and evolution
 of giant planets: The effects of solid cores. \icarus 67, 391-408.
\bibitem[Chambers(2006a)]{Chambers2006a} Chambers, J. E., 2006a.
 A semi-analytic model for oligarchic growth. \icarus 180, 496-513.
\bibitem[Chambers(2006b)]{Chambers2006b} Chambers, J. E., 2006b.
 Planet formation with migration. \apj 652, L133-L136.
\bibitem[Chambers(2008)]{Chambers2008} Chambers, J. E., 2008.
 Oligarchic growth with migration and fragmentation. \icarus 198, 256-273.
\bibitem[\'{C}uk and Burns(2004)]{Cuk2004} \'{C}uk, M., Burns, J. A., 2004.
 Gas-drag-assisted capture of Himalia's family. \icarus 167, 369-381.
\bibitem[D'Angelo et al.(2002)]{DAngelo2002} D'Angelo, G., Henning. T.,
 Kley, W., 2002. Nested-grid calculations of disk-planet interaction.
 \aap 385, 647-670.
\bibitem[Dones and Tremaine(1993)]{Dones1993} Dones, L., Tremaine,
 S., 1993.  On the origin of planetary spins. \icarus 103, 67-92.
\bibitem[Fortney et al.(2007)]{Fortney2007} Fortney, J. J., Marley,
 M. S., Barnes, J. W., 2007, Planetary radii across five orders of
 magnitude in mass and stellar insolation: application to transits.
 \apj 659, 1661--1672
\bibitem[Greenzweig and Lissauer (1990)]{GL90} Greenzweig, Y., Lissauer,
 J. J., 1990.  Accretion rates of protoplanets.  \icarus 89, 40-77.
\bibitem[Greenzweig and Lissauer (1992)]{GL92} Greenzweig, Y.,
 Lissauer, J. J., 1992.  Accretion rates of protoplanets II. Gaussian
 distributions of planetesimal velocities. \icarus 100, 440-463.
\bibitem[Hubickyj et al.(2005)]{Hubickyj2005} Hubickyj, O., Bodenheimer,
 P, Lissauer, J. J., 2005. Accretion of the gaseous envelope of
 Jupiter around a 5--10 Earth-mass core. \icarus 179, 415-431.
\bibitem[Ida (1990)]{Ida1990} Ida, S., 1990. Stirring and dynamical
 friction rates of planetesimals in the solar gravitational field.
 \icarus 88, 129-145.
\bibitem[Ida and Makino (1992)]{IM92} Ida, S., Makino, J., 1992.
 $N$-body simulation of gravitational interaction between planetesimals
 and a protoplanet. \icarus 96, 107-120.
\bibitem[Ida and Nakazawa (1989)]{IN89} Ida, S., Nakazawa, K., 1989.
 Collisional probability of planetesimals revolving in
 the solar gravitational field. III. \aap 224, 303-315.
\bibitem[Ikoma et al.(2000)]{Ikoma2000} Ikoma, M., Nakazawa, K.,
 Emori, H., 2000. Formation of giant planets: Dependences on
 core accretion rate and grain opacity. \apj 537, 1013-1025.
\bibitem[Inaba and Ikoma(2003)]{II03} Inaba, S., Ikoma, M., 2003.
 Enhanced collisional growth of a protoplanet that has an
 atmosphere. \aap 410, 711-723.
\bibitem[Inaba et al.(2001)]{Inaba2001} Inaba, S., Tanaka, H., Nakazawa,
 K., Wetherill, G. W., Kokubo, E., 2001. High-accuracy statistical
 simulation of planetary accretion: II. comparison with $N$-body
 simulation. \icarus 149, 235-250.
\bibitem[Inaba et al.(2003)]{Inaba_etal_2003} Inaba, S., Wetherill,
 G. W., Ikoma, M., 2003. Formation of gas giant planets: core
 accretion models with fragmentation and planetary envelope. \icarus
 166, 46-62.
\bibitem[Kary and Lissauer(1995)]{Kary1995} Kary, D. M., Lissauer,
 J. J., 1995.  Nebular gas drag and planetary accretion. II. Planet on
 an eccentric orbit. \icarus 117, 1-24.
\bibitem[Kary et al.(1993)]{Kary1993} Kary, D. M., Lissauer, J. J.,
 Greenzweig, Y., 1993.  Nebular gas drag and planetary accretion. 106,
 288-307.
\bibitem[Kokubo and Ida(1996)]{KokuboIda1996} Kokubo, E., Ida, S.,
 1996.  Oligarchic growth of protoplanets. \icarus 131, 171-178.
\bibitem[Kokubo and Ida(2000)]{KokuboIda2000} Kokubo, E., Ida, S.,
 2000.  Formation of protoplanets from planetesimals in the solar
 nebula. \icarus 143, 15-27.
\bibitem[Korycansky and Pollack(1993)]{Korycansky1993} Korycansky,
 D. G., Pollack, J. B., 1993. Numerical calculations of the linear
 response of a gaseous disk to a protoplanet. \icarus 102, 150-165.
\bibitem[Masset et al.(2006)]{Masset2006} Masset, F., D'Angelo, G.,
 Kley, W., 2006. On the migration of protogiant solid cores.
 \apj 652, 730-745.
\bibitem[Mizuno(1980)]{Mizuno1980} Mizuno, H., 1980. Formation of the
 giant planets. Prog. Theor. Phys. 64, 544-557.
\bibitem[Movshovitz and Podolak(2008)]{Movshovitz2008} Movshovitz, N.,
 Podolak, M., 2008.  The opacity of grains in protoplanetary
 atmospheres. \icarus 194, 368-378.
\bibitem[Nakazawa et al.(1989)]{Nakazawa1989} Nakazawa, K., Ida, S.,
 Nakagawa, Y., 1989. Collisional probability of planetesimals
 revolving in the solar gravitational field. I - Basic formulation.
 \aap 220, 293-300.
\bibitem[Nishida(1983)]{Nishida1983} Nishida, S., 1983.  Collisional
 processes of planetesimals with a protoplanet under the gravity of
 the proto-sun. Prog. Theor. Phys. 70, 93-105.
\bibitem[Ohtsuki(1993)]{Ohtsuki1993} Ohtsuki, K., 1993.  Capture
 probability of colliding planetesimals: Dynamical constraints on
 accretion of planets, satellites, and ring particles. \icarus 106,
 228-246.
\bibitem[Ohtsuki(1999)]{Ohtsuki1999} Ohtsuki, K., 1999.  Evolution of
 particle velocity dispersion in a circumplanetary disk due to inelastic
 collisions and gravitational interactions. \icarus 137, 152-177.
\bibitem[Ohtsuki et al.(2002)]{Ohtsuki2002} Ohtsuki, K., Stewart, G. R.,
 Ida, S., 2002.  Evolution of planetesimal velocities based on
 three-body orbital integrations and growth of protoplanets. \icarus
 155, 436-453.
\bibitem[Podolak et al.(1988)]{Podolak1988} Podolak, M., Pollack, J. B.,
 Raynolds, R. T., 1988.  Interactions of planetesimals with
 protoplanetary atmospheres. \icarus 73, 163-179.
\bibitem[Pollack et al.(1979)]{Pollack1979} Pollack, J. B., Burns,
 J. A., Tauber, M. E., 1979.  Gas drag in primordial circumplanetary
 envelopes -- A mechanism for satellite capture. \icarus 37, 587-611.
\bibitem[Pollack et al.(1996)]{Pollack1996} Pollack, J. B., Hubickyj,
 O., Bodenheimer, P, Lissauer, J. J., Podolak, M., Greenzweig, Y.,
 1996.  Formation of the giant planets by concurrent accretion
 of solids and gas. \icarus 124, 62-85.
\bibitem[Rafikov(2006)]{Rafikov2006} Rafikov, R. R., 2006.  Atmospheres
 of protoplanetary cores: Critical mass for nucleated instability.
 \apj 648, 666-682.
\bibitem[Sasaki and Nakazawa(1990)]{Sasaki1990} Sasaki, S.,
 Nakazawa, K., 1990.  Did a primary solar-type atmosphere exist around
 the proto-earth?  \icarus 85, 21-42.
\bibitem[Stevenson(1982)]{Stevenson1982} Stevenson, D. J., 1982.
 Formation of the giant planets. \planss, 30, 755-764.
\bibitem[Tanaka and Ida(1999)]{Tanaka1999} Tanaka, H., Ida, S.,
 1999.  Growth of a migrating protoplanet. \icarus 139, 350-366.
\bibitem[Tanaka et al.(2002)]{Tanaka2002} Tanaka, H., Takeuchi, T.,
 Ward, W. R., 2002.  Three-dimensional interaction between a planet and
 an isothermal gaseous disk. I. Corotation and Lindblad torques and planet
 migration. \apj 565, 1257-1274.
\bibitem[Ward(1986)]{Ward1986} Ward, W. R., 1986.  Density waves in the
 solar nebula - Differential Lindblad torque. \icarus 67, 164-180.
\bibitem[Ward(1997)]{Ward1997} Ward, W. R., 1997.  Protoplanet migration
 by nebular tides. \icarus 126, 261-281.
\bibitem[Wetherill and Cox(1985)]{Wetherill1985} Wetherill, G. W.,
 Cox, L. P., 1985.  The range of validity of the two-body approximation
 in models of terrestrial planet accumulation. II -
 Gravitational cross sections and runaway accretion. \icarus 63, 290-303.
\bibitem[Wetherill and Stewart(1989)]{WS89} Wetherill, G. W.,
 Stewart, G. R., 1989.  Accumulation of a swarm of small planetesimals.
 \icarus 77, 330-357.

\end{thebibliography}
\end{document}